\PassOptionsToPackage{pdfpagelabels=false}{hyperref} 
\documentclass[fleqn,usenatbib]{mnrasC}

\usepackage[T1]{fontenc}
\usepackage{ae,aecompl}
\usepackage{advdate}
\usepackage{siunitx}
\usepackage{color}
\usepackage{graphicx}	
\usepackage{amsmath}	
\usepackage{amssymb}	
\usepackage{hyperref}

\usepackage{txfonts}

\definecolor{UniRot}{rgb}{0.65,0.12,0.22} 
\definecolor{BeamerGruen}{RGB}{0,96,0} 
\definecolor{PraesBlau}{RGB}{38,38,134} 
\definecolor{OoGruen}{RGB}{87,157,28} 

\DeclareMathOperator{\e}{e}

\DeclareMathOperator{\const}{const}

\renewcommand{\vec}[1]{\boldsymbol{#1}}
 
\renewcommand{\d}{{\rm d}}
\renewcommand*\partial{\textsf{\reflectbox{6}}}

\title[Stability Analysis of Magnetised Neutron Stars]{Stability Analysis of Magnetised Neutron Stars - A Semi-analytic Approach}

\author[Marlene Herbrik, Kostas D. Kokkotas]{
Marlene Herbrik$^{1}$\thanks{E-mail: marlene.herbrik@uni-tuebingen.de},
Kostas Kokkotas$^{1}$
\\
$^{1}$Theoretical Astrophysics, IAAT, Eberhard Karls University of T{\"u}bingen, 72076 T{\"u}bingen, Germany\\
}

\date{\AdvanceDate[-1]\today}

\pubyear{2015}

\begin{document}
\label{firstpage}
\pagerange{\pageref{firstpage}--\pageref{lastpage}}
\maketitle

\begin{abstract}
 We implement a semi-analytic approach for stability analysis, addressing the ongoing uncertainty about stability and structure of neutron star magnetic fields. Applying the energy variational principle, a model system is displaced from its equilibrium state. The related energy density variation is set up analytically, whereas its volume integration is carried out numerically. This facilitates the consideration of more realistic neutron star characteristics within the model compared to analytical treatments. At the same time, our method retains the possibility to yield general information about neutron star magnetic field and composition structures that are likely to be stable. In contrast to numerical studies, classes of parametrized systems can be studied at once, finally constraining realistic configurations for interior neutron star magnetic fields. We apply the stability analysis scheme on polytropic and non-barotropic neutron stars with toroidal, poloidal and mixed fields testing their stability in a Newtonian framework. Furthermore, we provide the analytical scheme for dropping the Cowling approximation in an axisymmetric system and investigate its impact. Our results confirm the instability of simple magnetised neutron star models as well as a stabilisation tendency in the case of mixed fields and stratification. These findings agree with analytical studies whose spectrum of model systems we extend by lifting former simplifications.
\end{abstract}



\section{Introduction}

Neutron stars represent the objects possessing the strongest magnetic fields known to appear in the universe. Typically, they show surface magnetic fields of \SIrange[range-units=single]{e11}{e13}{G} as inferred by spin-down rates of pulsar observations \citep{HewishBell, Gold}. Furthermore, bursts, QPOs and spin-down rates provide observational evidence for even stronger fields of \SIrange[range-units=single]{e13}{e15}{G} in the exterior and interior of magnetars \citep{DuncanThompson, ThompsonDuncan}. Interior fields are generally expected to be stronger than observably accessible surface fields since there are indications that exterior poloidal magnetic fields are stabilised by interior toroidal components exceeding a minimum strength \citep{Tayler1, LanderJones2012, Akguen}. According to this, classical pulsars may contain a toroidal field of the strength of \SIrange[range-units=single]{e12}{e14}{G}.\par
These field strengths exceed magnetic field intensities producible on earth by many orders of magnitude. Providing such long-lastingly high field strengths, neutron stars are unique laboratories for gaining insights about properties of matter under extreme conditions. At the same time, magnetic fields in magnetars are strong enough to have an impact on fundamental neutron star processes, manifested in the equation of state, surface composition and geometry of the star. A detailed knowledge about the interior magnetic field structure is therefore highly desired and inevitable for many fields of neutron star physics. It is an essential ingredient for the interpretation of cooling curves, simulation of neutron star binary mergers as well as neutron star mode analysis and the attempt of constraining the equation of state using neutron star seismology. Nevertheless, the detailed interior field configuration is still widely unknown, representing an unresolved key issue of astrophysics. So far, there exist no generic magnetic field equilibria that produce intrinsically stable neutron star models, although this problem has been addressed in various studies with different levels of model complexity.\par\bigskip

According to simple theoretical models, magnetised neutron stars are unstable. More realistic models complicate an analytical treatment considerably. Numerical attempts on the other hand are not able to provide general conclusions about classes of systems.\par
Analytical studies have shown that neutron stars with purely toroidal and purely poloidal magnetic fields are unstable \citep{Wright, Tayler1, Tayler2, Tayler3}, suffering pinch instabilities of a cylindrical fluid discharge (hereafter referred to as Tayler instabilities). Mixed fields and stratification show indications of having a stabilising impact on the system \citep{Tayler4, Akguen}. These studies are based on the energy variational principle, where stability/instability of the system can be inferred from the positive/negative sign of the change in the total system energy during a displacement. Partly strong simplifications are necessary to allow for an analytical treatment. They prohibit us from gaining the general information being necessary in order to construct a complete model for magnetised neutron stars. Common assumptions for stability analysis are a polytropic equation of state, spherical symmetry, specified magnetic field, composition and perturbation structures as well as neglecting the change in the gravitational potential caused by the displacement (hereafter referred to as Cowling approximation).\par
Numerical simulations on the other hand suggest the so called twisted torus field to be stable on time-scales long compared to the Alfv\'{e}n time \citep{BraithwaiteNordlund, BraithwaiteSpruit}. Although numerical studies \citep{Braithwaite, BraithwaiteA, Colaiuda, Ciolfi, LanderJones2010A, LanderJones2010B, Lasky}, don't rely on simplifications and recently also considered mixed fields and stratification \citep{BraithwaiteB, LanderJones2012}, they require a precise specification of initial conditions. Therefore, they are not a convenient tool for constraining realistic field structures out of a variety of possibilities.\par\bigskip

This difficulty motivates the idea of a semi-analytic approach for stability analysis implemented in this work.\par
We will test stability according to the energy variational principle by calculating the change in the total system energy during a displacement. The energy variational density will be set up \emph{analytically}, thus preserving generality and complexity of the studied system. Arbitrary system characteristics such as composition and magnetic field structure are parametrized. Subsequently we integrate the variational density \emph{numerically} in order to obtain the energy variation of the chosen model. That way, we avoid the hitherto necessary simplifications. With our method, it is affordable to consider mixed fields and stratification in the star and drop Cowling's approximation which has been commonly used in former studies.\par
Beginning with a simple system known to be unstable, additional more realistic model features can be taken into account, investigating their potential stabilising impact. We can calculate the energy variation and test its sign for varying sets of parameters. This method is able to reveal stellar characteristics that are more likely to result in a positive energy variation and thus stable model configuration than others. That way, realistic magnetic field and composition structures for neutron stars can be constrained.\par
Another advantage of this semi-analytic approach lies in facilitating the search for stable equilibria, currently becoming more relevant. Unlike earlier studies that focussed on \emph{finding instabilities} in simple models \citep{Wright, Tayler1, Tayler2, Tayler3}, recent works such as \citet{BraithwaiteA, BraithwaiteB, LanderJones2012, Akguen} consider the question of \emph{how to explain} the actual \emph{stability} of observed neutron stars. More precisely, they try to find out which features of a more realistic star are responsible for its stability. With this attempt, new challenges come along: First, the Cowling approximation, possibly tending to let a system appear more stable than it actually is, might not be valid any more. Second, finding stability involves exposing a system to all possible perturbations, whereas an instability proof only requires finding one perturbation causing an instability. We show how both issues can be addressed in the semi-analytic approach, enabled by the advantages the numerical procedure has. Cowling's approximation can be dropped, possible perturbations can be expressed in a systematic way, 
without making the system more complex than treatable.\par\bigskip

The following section \ref{AnalyticModelling} describes the modelling of the system as well as the analytic setup of the energy variational density. In section \ref{NumericalProcedure} we explain how to solve the energy variational integral numerically. In section \ref{Applications} the scheme we developed is applied on magnetised neutron stars verifying Tayler's instabilities for toroidal and poloidal fields (sections \ref{PurelyToroidalMagneticField} and \ref{PurelyPoloidalMagneticField}) and showing the stabilising impact of stratification and mixed fields (section \ref{MixedFieldAndStratification}). We evaluate there as well, in which cases a full non-Cowling treatment is important for stability analysis. Finally, we conclude with section \ref{Conclusions} by discussing relevance and functionality of the semi-analytic approach and some of its numerous possible applications. Mathematical derivations and the description of the numerical scheme being used are presented in the appendix.

\section{Analytical modelling}
\label{AnalyticModelling}
\subsection{Model system}
\label{ModelSystem}
We consider an axisymmetric magnetised neutron star without rotation as an equilibrium configuration. The neutron star is perturbed by an arbitrary displacement field. Equilibrium quantities are denoted by the index $0$. Due to the structure of the magnetic field, axisymmetry is an appropriate choice reducing the spatial dimensions to two. Neglecting rigid rotation is a simplification not having a major impact on stabilising magnetic instabilities as shown by \citet{Tayler6, LanderJones2012}. As long as the equilibrium system is only slightly magnetically deformed, it can approximately be described as a non-magnetised spherically symmetric background. We work in geometrised Gaussian units.

\subsubsection{Coordinate systems being used}
\label{CoordinateSystemsBeingUsed}
As we work with various magnetic field structures and compare our results to different studies, we use three kinds of coordinate systems, adjusting them optimally to the problems under consideration.\par
\begin{figure}
	\includegraphics[width=0.48\columnwidth]{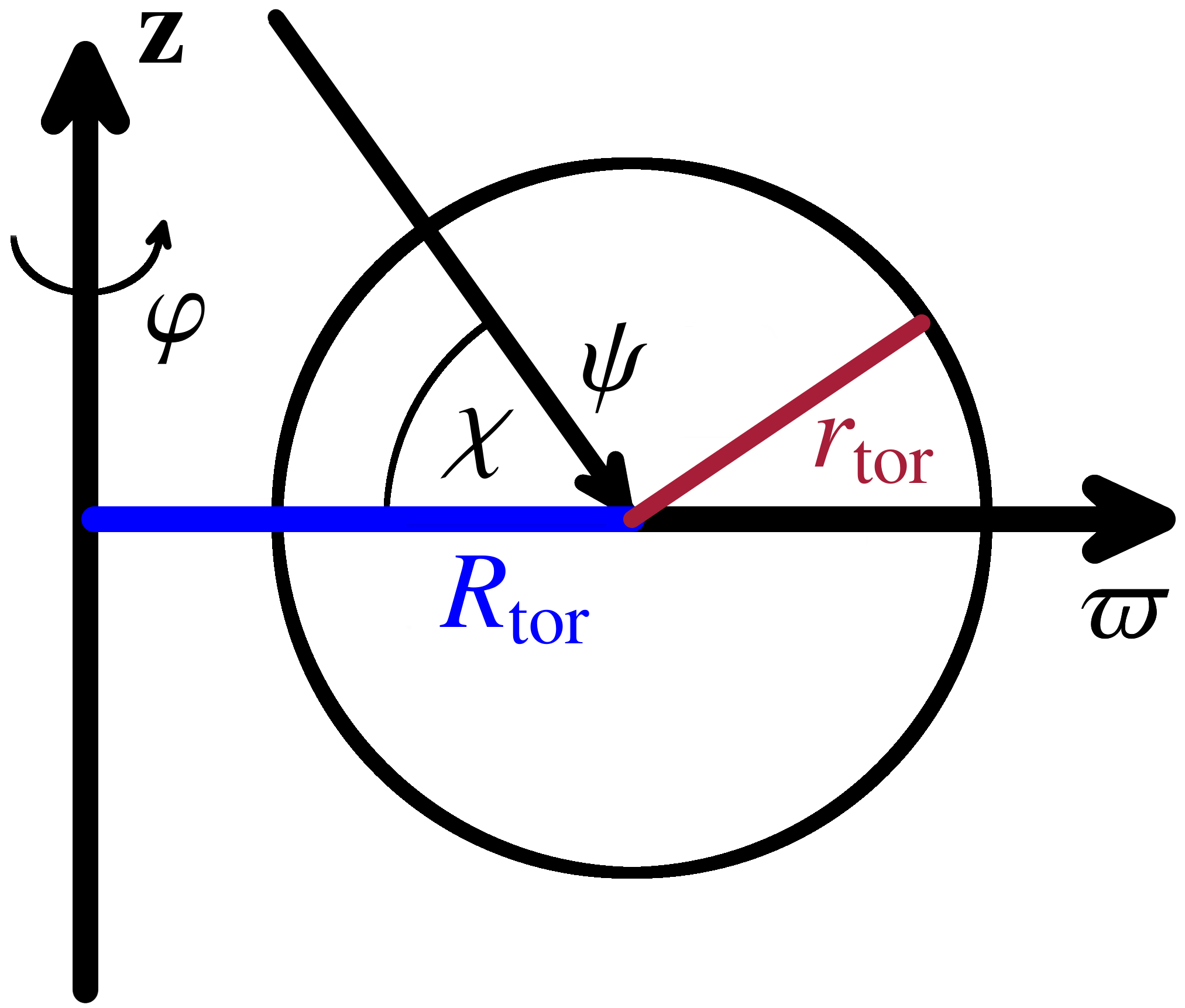}
	\includegraphics[width=0.48\columnwidth]{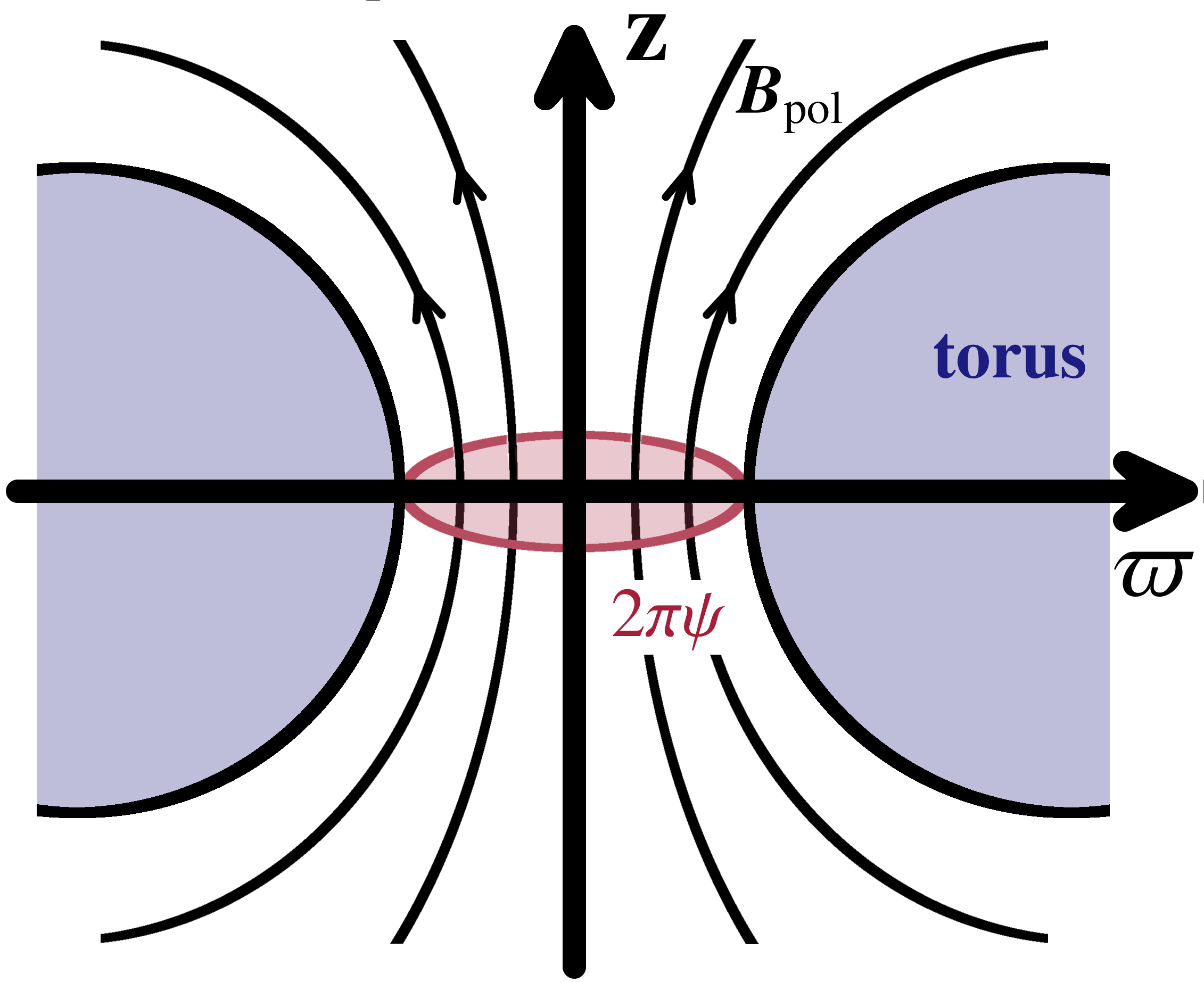}
    \caption{Toroidal coordinates $(\psi, \varphi, \chi)$ (left). The magnetic flux of a poloidal field along the stellar symmetry axis -- enclosed by symmetry axis and torus surface -- is determined as $2 \pi \psi$ (right).}
\label{fig:ToroidalCoordinates}
\end{figure}

For a treatment of the spherically symmetric unmagnetised background and comparison with work by \citet{Akguen}, spherical coordinates $(r,\vartheta,\varphi)$ are used. In presence of the magnetic field, cylindrical coordinates $(\varpi, \varphi, z)$ are the most convenient choice. If the field is purely poloidal, a toroidal coordinate system $(\psi, \varphi, \chi)$ as proposed by \citet{Tayler2} and shown in Fig.~\ref{fig:ToroidalCoordinates} fits best. The transformation equations between cylindrical and toroidal coordinates are
\begin{equation}
\label{equ:ToroidalCoordinates}
\begin{array}{rl}
          \psi &= - \frac{1}{2} B_\chi \bar r\\
          \varphi &= \varphi\\
           \chi &= \arctan{\frac{z}{R_{\text{tor}}-\varpi}}
         \end{array}
          \qquad\qquad
 \begin{array}{rl}
          \varpi &= R_{\text{tor}}-\bar r \cos \chi\\
          \varphi &= \varphi\\
          z &= \bar r \sin \chi.
         \end{array}       
\end{equation}
$\varphi$ is the same azimuthal angle as in cylindrical and spherical coordinates. $\psi$ and $\chi$ act as polar coordinates in the $(\varpi, z)$-plane on a torus cross section. $R_{\text{tor}}$ denotes the distance of the torus centre from the stellar symmetry axis in the equatorial plane.
\begin{equation}
\bar r= \sqrt{\left(R_{\text{tor}}-\varpi\right)^2 + z^2} =\sqrt{-\frac{2 \psi}{B_\chi}}
\end{equation}
measures the distance of a point from the torus centre. Note that the unit vector $\vec e_{\psi}$ points towards the torus centre with $\psi<0$, contrary to typical definitions of polar coordinates. The coordinate $\psi$ acts as a stream function defining the axisymmetric poloidal field $\vec B_{\text{pol}} = (0, 0, B_\chi)$.

\subsubsection{Unmagnetised equilibrium background}
We assume a two-component fluid consisting of a neutral fluid of neutrons $n$ and a charged one of protons $p^+$ and electrons $e^-$, both being in $\beta$-equilibrium. Hence, at every position $r$ within the star, the ratio of proton and neutron mass densities, $x^p=\rho_p/\rho_n$, takes the energetically most favourable value. It is determined by the weak force driven $\beta^-$- and inverse $\beta^-$-decay reactions
\begin{subequations}
\begin{align}
n &\longrightarrow p^+ + e^- + \bar\nu_e\\
p^+ + e^- &\longrightarrow n + \nu_e,
\end{align}
\end{subequations}
where $\nu_e$ and $\bar\nu_e$ denote electron neutrino and anti-neutrino. In spherical coordinates $(r, \vartheta, \varphi)$, the spherically symmetric equilibrium system is basically one-dimensional. The radial distribution of enclosed mass $m_0$, pressure $p_0$ and gravitational potential $\Phi_0$ in the unmagnetised background is given by the Newtonian system equations \citep{Shapiro}
\begin{subequations}
\label{equ:SystemEquations}
\begin{align}
 d_r m_0 &=4 \pi r^2 \rho_0\\
\label{equ:SystemEquationsP}
d_r p_0 &= -\frac{\rho_0 m_0}{r^2}\\
\label{equ:SystemEquationsPhi}
d_r \Phi_0 &= -\frac{1}{\rho} d_r p_0,
\end{align}
\end{subequations}
with $\rho_0$ denoting the total mass density. The system is closed by a polytropic equation of state \eqref{equ:EquationOfStateBackground}. Relations \eqref{equ:SystemEquationsP} and \eqref{equ:SystemEquationsPhi} follow from hydrostatic equilibrium equation \eqref{equ:HydrostaticEquilibrium} and Poisson equation \eqref{equ:PoissonEquation} in their general forms:
\begin{align}
\label{equ:EquationOfStateBackground}
 p_0 &= \kappa \rho_0^{\Gamma_0}\\
\label{equ:HydrostaticEquilibrium}
\vec \nabla p_0 &= \pm \rho_0 \vec \nabla \Phi_0\\
\label{equ:PoissonEquation}
\vec \nabla^2 \Phi_0 &= 4 \pi \rho_0.
\end{align}
$\Gamma_0$ is the polytropic exponent for the background system and $\kappa$ a proportionality constant. Upper and lower sign correspond to the definition of the gravitational field $\vec g$ parallel or antiparallel to the gradient of the gravitational potential:
\begin{equation}
\label{equ:GravitationalField}
 \vec g = \pm \vec \nabla \Phi.
\end{equation}
Assuming a central density $\rho_c$, $\kappa$ and $\Gamma_0$, a neutron star model with total mass $M$ and radius $R$ is fully determined.

\subsubsection{Magnetic field}
\label{MagneticField}
\begin{figure}
	\includegraphics[width=\columnwidth]{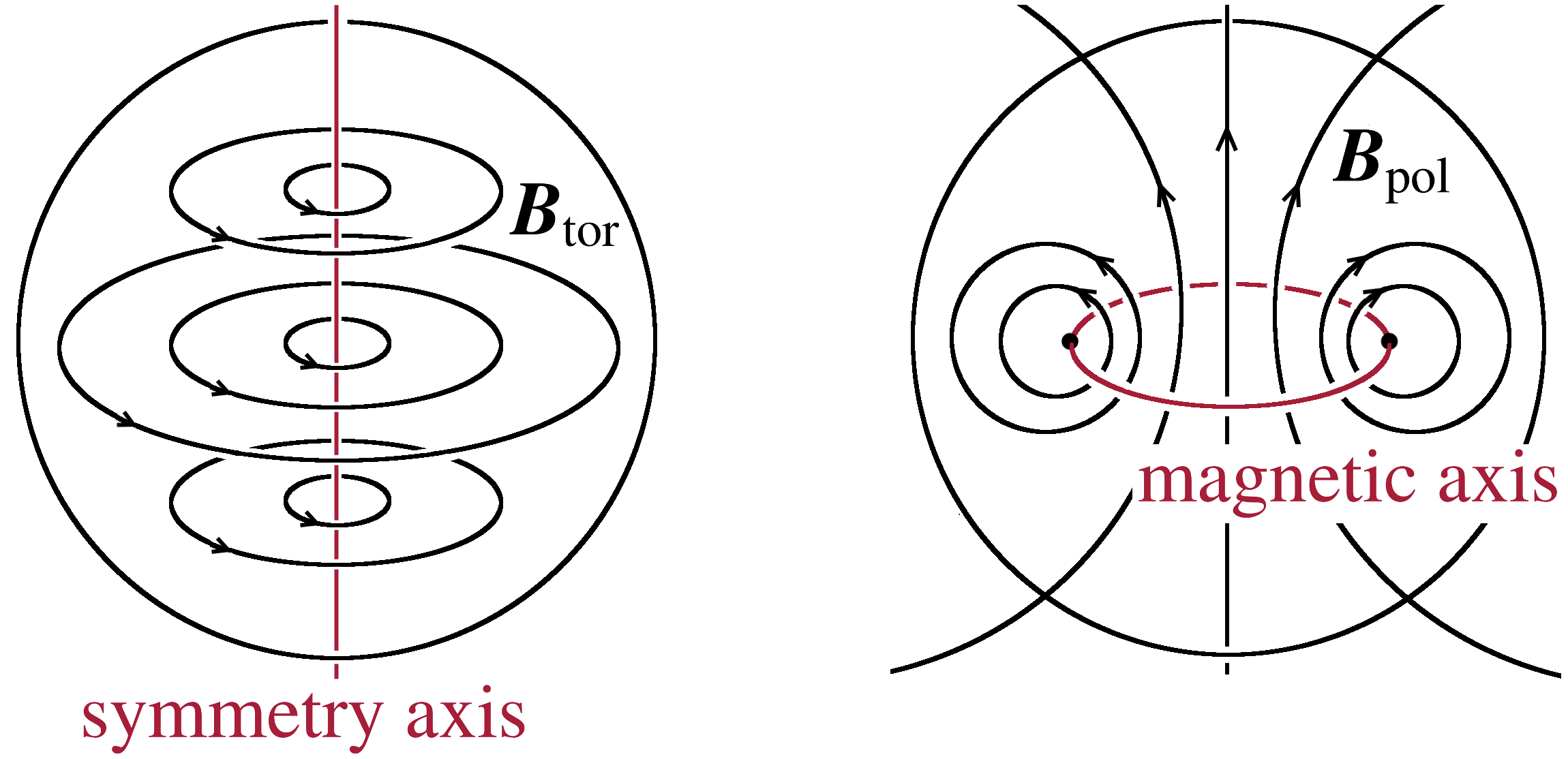}
    \caption{Field lines and symmetry axes of toroidal (left) and poloidal (right) magnetic field components.}
\label{fig:TorPol}
\end{figure}

Consisting of conducting particles, the charged fluid evolves according to ideal magnetohydrodynamics \citep{Mestel,Shapiro,Thompson} with the assumption
\begin{equation}
\label{equ:ForceFreeCondition}
 \vec E + \vec v \times \vec B \approx 0
\end{equation}
of an electromagnetically force-free medium for the electric field $\vec E$, particle velocity $\vec v$ and magnetic field $\vec B$. The basic equations for the magnetised system are the Euler equation of motion \eqref{equ:EulerEquation}, Amp\`{e}re's law, the equation of magneto-kinematics \eqref{equ:Magnetokinematics} and the continuity equation of mass conservation \eqref{equ:ContinuityEquationEuler}:
\begin{align}
\label{equ:EulerEquation}
\rho \d_t\vec v &= - \vec \nabla p \pm \rho \vec \nabla \Phi + q\vec j \times \vec B\\
\label{equ:AmperesLaw}
\vec \nabla\times \vec B &= 4 \pi \vec j\\
\label{equ:Magnetokinematics}
 \partial_t \vec B &= \vec \nabla \times \left( \vec v \times \vec B \right)\\
\label{equ:ContinuityEquationEuler}
 \partial_t \rho &= - \vec \nabla\cdot \left(\rho \vec v \right),
\end{align}
with $\partial$ denoting partial, $\d$ denoting full derivatives. $\vec j$ is the current density creating the magnetic field. Due to axisymmetry, the system is basically two-dimensional in cylindrical coordinates $(\varpi, \varphi, z)$ and every possible magnetic field 
\begin{equation}
 \vec B = \vec B_{\text{tor}} + \vec B_{\text{pol}}
\end{equation}
can be expressed as the sum of a toroidal and poloidal field component \citep{GradRubin}. See Fig.~\ref{fig:TorPol} for the field line structure of these components.
\par\bigskip
The semi-analytic approach is able to treat arbitrary magnetic field structures. In our first applications of the scheme (see section \ref{Applications}), we use the following ones.\par
\begin{figure}
	\includegraphics[width=\columnwidth]{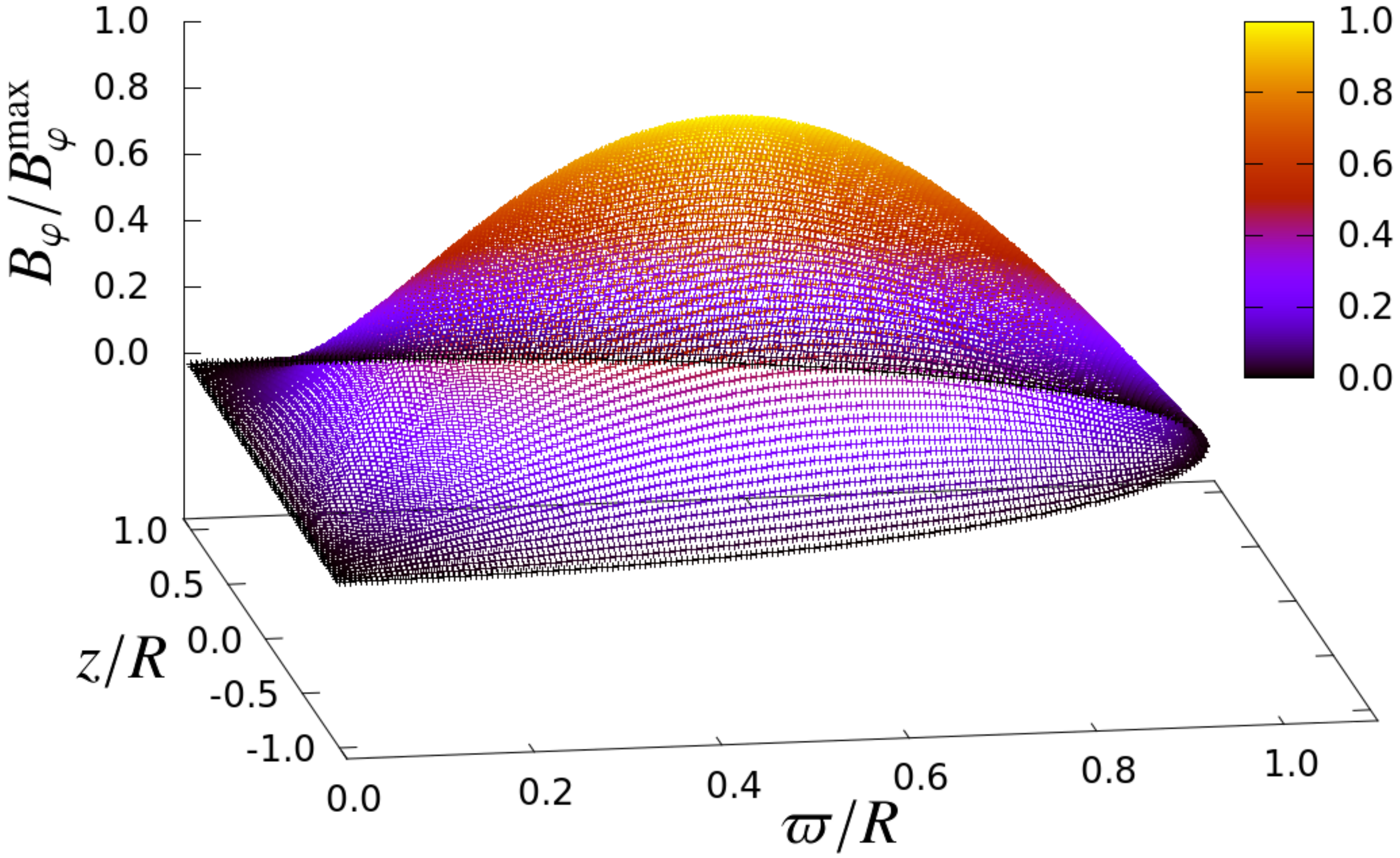}
    \caption{Spatial distribution of the toroidal field strength $B_\varphi$ defined by equations \eqref{equ:BTaylerToroidal}.}
\label{fig:Btor}
\end{figure}

First, we assume \emph{purely toroidal} magnetic fields
\begin{subequations}
 \label{equ:BTaylerToroidal}
\begin{align}
 B_\varpi&=0\\
 B_\varphi &= \varpi \rho_0 B_{\text{tor}}\\
 B_z &= 0
\end{align}
\end{subequations}
according to \citet{Tayler1}. The explicit form of $B_\varphi$ in \eqref{equ:BTaylerToroidal} lets the magnetic field vanish at the symmetry axis, $\varpi=0$, and the surface, where $\rho_0=0$, see Fig.~\ref{fig:Btor}. $B_{\text{tor}}$ is a dimensionless constant measuring the strength of the toroidal field.\par
Second, \emph{purely poloidal} magnetic fields
\begin{subequations}
 \label{equ:BTaylerPoloidalZwischenschritt}
\begin{align}
 B_\varpi&=-\frac{1}{\varpi} \partial_z \psi\\
 B_\varphi &= 0\\
 B_z &= \frac{1}{\varpi} \partial_\varpi \psi
\end{align}
\end{subequations}
are assumed according to \citet{Tayler2}. $\vec B$ can be expressed in terms of a stream function $\psi$ \citep{GradRubin, LanderJones2009}. In toroidal coordinates defined in \eqref{equ:ToroidalCoordinates}, the poloidal field \eqref{equ:BTaylerPoloidalZwischenschritt} can be written as
\begin{subequations}
 \label{equ:BTaylerPoloidal}
\begin{align}
 B_\psi&=0\\
 B_\varphi &= 0\\
 B_\chi &=\frac{r_{\text{tor}}}{\varpi} B_{\text{pol}}.
\end{align}
\end{subequations}
The poloidal field vanishes at its symmetry axis which is a circular line around the stellar symmetry axis in the equatorial plane and will be referred to as magnetic axis, see Fig.~\ref{fig:TorPol}. The particular choice of $B_\chi$ is linked to the definition of the toroidal coordinates: As illustrated in Fig.~\ref{fig:ToroidalCoordinates}, $2 \pi \psi$ is the magnetic flux along $z$ between symmetry axis and the torus on which $\psi$ and $\chi$ are defined.  Following Tayler's choice, $B_{\text{pol}} [\si{cm^{-2}}]$ is a constant measuring the strength of the poloidal field. In general, however, $B_{\text{pol}} (\psi)$ contains the dependency of $\vec B$ on the stream function shown in expression \eqref{equ:BTaylerPoloidalZwischenschritt}.\par
Investigating \emph{mixed fields with stratification}, we follow \citet{Akguen} assuming
\begin{subequations}
\label{equ:BAkguen}
\begin{align}
 \vec B_{\text{tor}} &= B_0 \eta_{\text{tor}} R  \hat \beta(r,\vartheta) \vec \nabla_r \varphi\\
 \vec B_{\text{pol}} &= B_0 \eta_{\text{pol}} R^2 \vec \nabla_r  \hat\alpha(r,\vartheta) \times \vec \nabla_r \varphi
\end{align}
\end{subequations}
in spherical coordinates with
\begin{equation}
\hat \alpha(r,\vartheta) = f(x) \sin^2 \vartheta, \qquad  \hat\beta(\varpi, z) = \begin{cases}
										    (\hat\alpha -1)^2 	&\text{ for } \hat \alpha \ge 1,\\
										    0	 		&\text{ for }  \hat\alpha < 1.
										    \end{cases}
\end{equation}
Let the radial function $f(x)$ with $x=r/R$ be
\begin{equation}
 f(x) = \frac{35}{8} x^2 -\frac{21}{4} x^4 + \frac{15}{8} x^6 \quad \text{for} \quad x \le 1
\end{equation}
inside the star. This choice fulfils a boundary condition at the stellar surface that allows for the exterior magnetic field to have a dipole structure. Furthermore, it obeys conditions in the centre of the star ensuring the magnetic field and its generating current density to be finite. $\vec \nabla_r$ takes derivatives with respect to $r$, not $x$. $\eta_{\text{tor}}$ and $\eta_{\text{pol}}$ are dimensionless prefactors and measure the maximum strengths of both field components via
\begin{align}
 B_{\text{tor}}^{\text{max}} &= B_{\text{tor}}(r\approx 0.782, \vartheta=\pi/2) \approx 0.0254 \eta_{\text{tor}} B_0\\
 B_{\text{pol}}^{\text{max}} &= B_{\text{pol}}(r=0, \vartheta=0) = \frac{35}{4} \eta_{\text{pol}} B_0,
\end{align}
where $B_0[\si{cm^{-2}}]$ is a constant amplitude. The toroidal field is nonzero only inside the region where poloidal field lines close inside the star (cf. Fig.~\ref{fig:TorPol}). The system is unstable for $B_{\text{pol}}=0$.

\subsubsection{Perturbed system}
During a perturbation, the position vector can be expressed by adding the displacement field $\vec \xi$ to the position vector $\vec r_0$ of the magnetised equilibrium star:
\begin{equation}
\label{equ:DefinitionXi}
 \vec r(t) = \vec r_0 + \vec \xi(\vec r,t).
\end{equation}
Consequently, all system quantities $Q$ in the perturbed state are given by
\begin{align}
 Q^E(\vec r,t) &= Q_0(\vec r) + \delta Q\\
 Q^L(\vec r,t) &= Q_0(\vec r) + \Delta^L Q.
\end{align}
Eulerian perturbations $\delta Q$ and Lagrangian perturbations $\Delta^L Q$ describe the deviation of $Q$ caused by $\vec \xi = \Delta^L \vec r$ from the equilibrium value $Q_0$ at a fixed position or a given fluid element, respectively.
\begin{equation}
\label{equ:DeltaQdeltaQ}
 \Delta^L Q = \delta Q + \vec \nabla_{\vec r_0} Q_0
\end{equation}
holds.

\subsubsection{Stratification}
\label{Stratification}
\begin{figure}
	\includegraphics[width=\columnwidth]{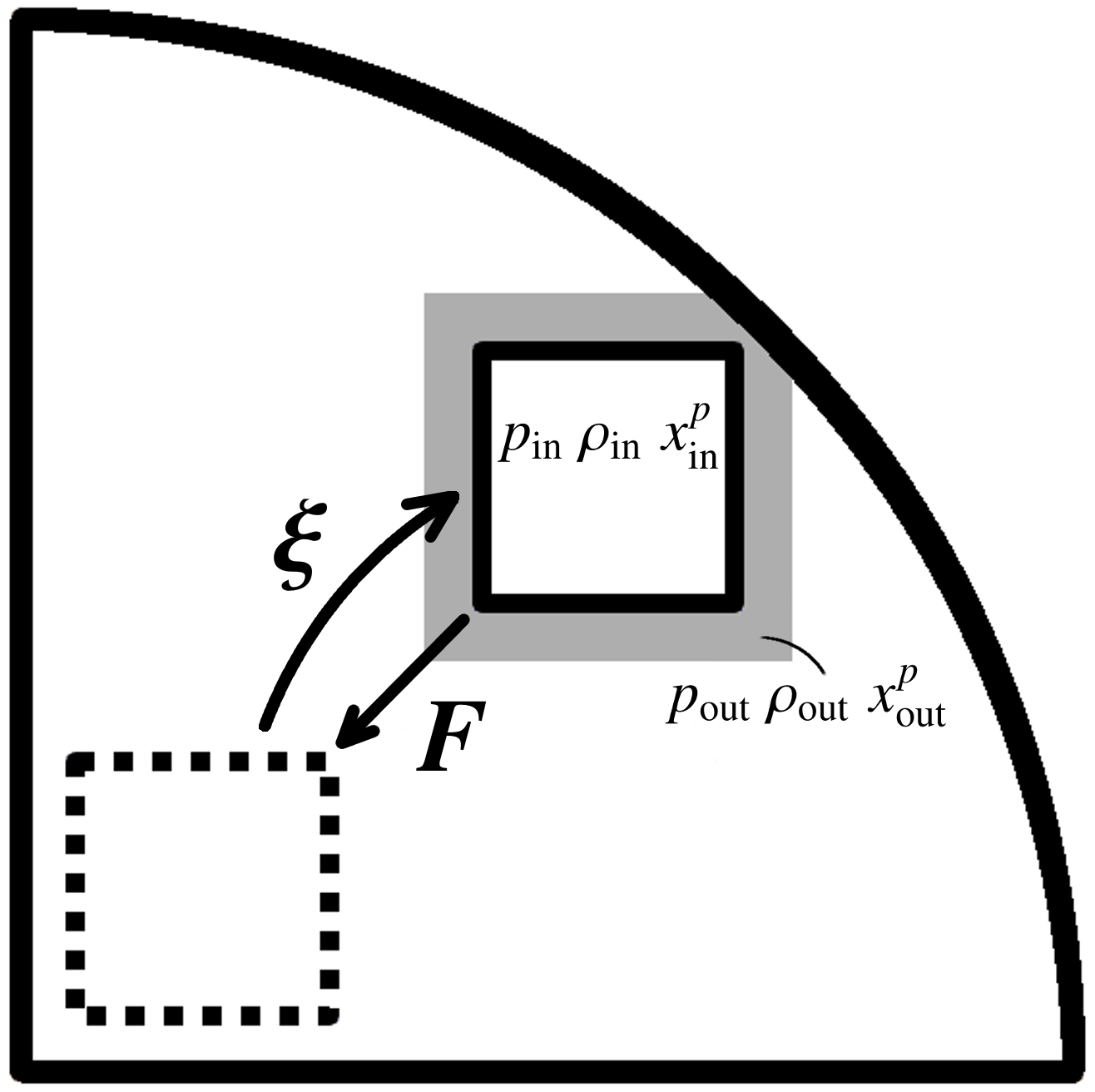}
    \caption{Stable g-mode type perturbation caused by stratification. A fluid element has been displaced radially outwards by $\vec \xi$. $p_{\text{in}}=p_{\text{out}}$ and $x_{\text{in}} \neq x^p_{\text{out}}$ hold. Due to $p=p(\rho, x^p)$, it is $\rho_{\text{in}} \neq \rho_{\text{out}}$. The difference in density causes a stratificationally driven gravitational force $\vec F$ on the fluid element.}
\label{fig:Stratification}
\end{figure}

In order to allow for the system being non-barotropic, the star is assumed to be stratified with $x^p(r) \neq \const$ throughout the star. Hence, pressure depends not only on density but also on proton fraction, $p=p(\rho, x^p)$. The impact of stratification on stability follows by introducing a polytropic exponent 
\begin{equation}
\label{equ:Gamma1}
 \Gamma_1 = \left( \frac{\partial \ln p_0}{\partial \ln \rho_0} \right)_{x^p}
\end{equation}
representing the change in $\ln p_0$ with $\ln \rho_0$ for fixed proton fraction \citep{ReiseneggerGoldreich, Passamontilang, Gaertig, Passamontikurz}. In a stratified star, it differs from
\begin{equation}
\label{equ:Gamma0}
 \Gamma_0 = \frac{\d \ln p_0}{\d \ln \rho_0} = \Gamma_1 + \left( \frac{\partial \ln p_0}{\partial \ln x_0^p} \right)_\rho \frac{\d \ln x_0^p}{\d \ln \rho_0},
\end{equation}
describing the change in $\ln p_0$ with $\ln \rho_0$ in $\beta$-equilibrium. Consequently, the equation of state for the perturbed quantities,
\begin{equation}
\label{equ:EquationOfStatePerturbations}
 \Delta^L p = \frac{p_0}{\rho_0} \Gamma_1 \Delta^L \rho,
\end{equation}
is different from the equilibrium equation of state \eqref{equ:EquationOfStateBackground}. Unlike in a barotropic star, a fluid element being displaced to a position with $x^p$ different from $x^p_0$, feels an additional gravitational force. While the fluid pressure adjusts quickly to the surroundings, the beta-reaction time-scale is longer than typical fluid oscillations, leaving the proton fraction inside the perturbed element constant, $\Delta^L x^p = 0$. The displaced element being too dense or less dense according to its new position, feels a force pushing it back to its original position or further away from it, corresponding to a stable vs. unstable g-mode \citep{Robe, Finn, Aerts}, illustrated in Fig.~\ref{fig:Stratification}. Direction and strength of the force are proportional to the sign and absolute value of $\Delta \Gamma \equiv \Gamma_1-\Gamma_0$.

\subsection{Energy variational principle}
\label{EnergyVariationalPrinciple}
\subsubsection{Stability criterion}
According to the energy variational principle, an equilibrium state is stable/unstable towards a perturbation mode $\vec \xi^n$ if it is energetically more/less favourable than the perturbed state. Thus, for a system in equilibrium the stability criterion
 \begin{equation}
 \label{equ:StabilityCriterion}
 \text{global } \left\{ \begin{array}{c} \text{stability}\\
					\text{instability}
                       \end{array} \right\}
 \Leftrightarrow \left\{ \begin{array}{ll}
                          \delta W >0  &\forall \quad \vec\xi^n\\
                          \delta W <0  &\text{for at least one } n
                         \end{array} \right\}
 \end{equation}
holds. $\delta W$ denotes the second variation of the total system energy caused by the displacement. Note that the first variation vanishes as the unperturbed system is assumed to be in equilibrium.\par
For the purpose of analysing the stability of neutron stars, we prefer the energy variational principle over normal mode analysis. Information about stability is easily accessible by checking the sign of the energy variation $\delta W$, not requiring the solution of eigenvalue equations. We are not interested in the time development of possible instabilities. Once a stable configuration has been found using the energy variational principle, time evolution can be investigated with simulation codes. The instabilities being found in magnetised neutron stars usually grow within a few Alfv\'{e}n crossing time-scales. Having such high growth rates, the located instabilities are dynamical and cannot be damped by any other mechanism.

\subsubsection{Energy variation}
In order to test stability via criterion \eqref{equ:StabilityCriterion}, the energy variation has to be calculated \citep{Bernstein, Mestel}. We outline the procedure here, for a detailed derivation see appendix \ref{DerivationEnergyVariation}. $\delta W$ can be derived perturbing the Euler equation \eqref{equ:EulerEquation}, yielding a sum of energy variational density contributions caused by magnetic field, fluid pressure and gravity, integrated over the volume:
\begin{small}
\begin{equation}
\label{equ:deltaW}
 \delta W = \iiint\limits_V \left[ \mathcal E_{\text{\text{magn}}} + \mathcal E_{\text{fluid}} + \mathcal E^{\text{Cowl}}_{\text{grav}} + \mathcal E^{\text{nC}}_{\text{grav}} \right] \d V.
\end{equation}
\end{small}
For calculating the total system energy variation, the integration area is equivalent to the stellar volume $V$. Smaller integration areas are possible if not all parts of the star contribute to $\delta W$ or the energy variation is defined for a sub-domain. Assuming the model system defined in section \ref{ModelSystem}, taking stratification and non-Cowling contribution $\delta W_{\text{grav}}^{\text{nC}}$ into account, the summands are
\begin{small}
\begin{subequations}
\label{equ:deltawiZwischenschritt}
\begin{align}
 \mathcal E_{\text{magn}} &= \frac{1}{2} \Re \left\{ \vec \xi^* \cdot \left[ \left(\delta \vec j \times \vec B_0 \right) + \left( \vec j_0 \times \delta \vec B \right) \right] \right\}\\
 \mathcal E_{\text{fluid}} &= \frac{1}{2} \Re \left\{ \vec \xi^* \cdot \vec \nabla \delta p \right\}\\
 \mathcal E^{\text{Cowl}}_{\text{grav}} &= \mp \frac{1}{2} \Re \left\{\delta \rho \vec \xi^* \cdot \vec \nabla \Phi_0 \right\}\\
 \mathcal E^{\text{nC}}_{\text{grav}} &= \mp \frac{1}{2} \Re\left\{ \rho_0 \vec\xi^* \cdot \vec \nabla \delta \Phi \right\},
\end{align}
\end{subequations}
\end{small}
including Euler variations of all system quantities. $\mathcal E_{\text{magn}}$ denotes the magnetic field contributions to the energy variational density, $\mathcal E_{\text{fluid}}$ summarises the contributions caused by fluid pressure. The total gravitational field contribution is $\mathcal E_{\text{grav}} = \mathcal E_{\text{grav}}^{\text{Cowl}} + \mathcal E_{\text{grav}}^{\text{nC}}$. The total energy variational density is the integrand of the energy integral and will be denoted by $\mathcal E =  \mathcal E_{\text{\text{magn}}} + \mathcal E_{\text{fluid}} + \mathcal E_{\text{grav}}$. Generally, $\vec \xi$  is a complex quantity, so real parts and complex conjugates in \eqref{equ:deltawiZwischenschritt} ensure $\mathcal E$ to be physically reasonable. Perturbing the basic equations \eqref{equ:AmperesLaw}, \eqref{equ:Magnetokinematics}, \eqref{equ:EquationOfStateBackground}, \eqref{equ:ContinuityEquationEuler} and \eqref{equ:PoissonEquation}, $\delta Q$ can be expressed in terms of $\vec \xi$ and equilibrium quantities as
\begin{small}
\begin{subequations}
\label{equ:deltaQ}
\begin{align}
\label{equ:deltaj}
 \delta \vec j &= \vec \nabla \times \vec {\delta B} +\mathcal{O} \left(\delta^2 \right)\\
 \label{equ:deltaB}
 \delta \vec B &= \vec \nabla \times \left(\vec \xi \times \vec B_0\right) +\mathcal{O} \left(\delta^2 \right)\\
 \label{equ:deltap}
 \delta p &= -\Gamma_1 p_0 \vec \nabla \vec \xi - \vec \xi \cdot \vec \nabla p_0 +\mathcal{O} \left(\delta^2 \right)\\
 \label{equ:deltarho}
 \delta \rho &=  - \vec \nabla \left( \rho_0 \vec \xi\right) +\mathcal{O} \left(\delta^2 \right)\\
 \label{equ:deltaPhi}
 \delta \Phi &= G \int \rho_0(\vec r') \vec \xi(\vec r') \cdot \vec \nabla \frac{1}{|\vec r - \vec r'|} \d V'  +\mathcal{O} \left(\delta^2 \right),
\end{align}
\end{subequations}
\end{small}
where we take into account only first order perturbation terms. That way, we receive the energy variation in second order after multiplying the relations \eqref{equ:deltaQ} for $\delta Q$ with $\mathcal O(\vec \xi \propto \delta)$ in the integrand \eqref{equ:deltawiZwischenschritt}:
\begin{small}
\begin{subequations}
\label{equ:deltawi}
\begin{align}
 \mathcal E_{\text{magn}} &= \frac{1}{2} \Re\left[ \vec Q^* \cdot \vec Q-\vec j \cdot \left(\vec Q^* \times \vec \xi \right) \right]\\
 \label{equ:deltawfluid}
 \mathcal E_{\text{fluid}} &= \frac{1}{2} \Re\left[ \Gamma_1 p_0 \left( \vec \nabla \cdot \vec \xi^* \right)\left( \vec \nabla \cdot \vec \xi \right) + \left( \vec \xi^* \cdot \vec \nabla p_0 \right) \vec \nabla \cdot \vec \xi \right] \\
 \label{equ:deltawgravCowl}
 \mathcal E^{\text{Cowl}}_{\text{grav}} &= \mp \frac{1}{2} \Re\left[\left( \vec \xi^* \cdot \vec \nabla \Phi_0 \right) \vec \nabla\cdot \left(\rho_0 \vec \xi\right) \right]\\
 \label{equ:deltawgravnC}
 \mathcal E^{\text{nC}}_{\text{grav}} &= \mp \frac{1}{2} \Re  \left[\rho_0 \vec\xi^* \cdot \vec \nabla \delta \Phi \right],
\end{align}
\end{subequations}
\end{small}
with $\vec Q = \vec \nabla\times \left(\vec \xi \times \vec B \right)$. The explicit form of $\delta \Phi$ depends on the form of the displacement field and will be given in section \ref{CowlingApproximation} for an explicit choice of $\vec \xi$.

\subsubsection{Displacement field}
\label{DisplacementField}
Setting up the energy variation defined in \eqref{equ:deltaW} with the summands given by \eqref{equ:deltawi}, requires a given background configuration $\left[p_0(r), \rho_0(r), \Phi_0(r)\right]$, an arbitrary choice for the equilibrium magnetic field $\vec B$ and a specified displacement field $\vec \xi$.
The background configuration can be constructed solving the system equations \eqref{equ:SystemEquations} with an equation of state \eqref{equ:EquationOfStateBackground}. $\vec B$ will be assumed according to the definitions in section \ref{MagneticField}. An assumption for $\vec \xi$ will be made as follows.\par
Depending on the mode against which we want to investigate the stability behaviour, we use three different classes of displacement fields, assigned with the choices of magnetic field structures defined in section \ref{MagneticField}. Note that arbitrary magnetic and displacement field configurations can be used in the semi-analytic approach thanks to the fact that constraints are removed when the integration is carried out numerically instead of analytically . For an investigation of possible instabilities though, the displacement field has to be chosen in accordance with the particular magnetic field structure in order to make instabilities visible that are potentially localised to certain regions of the field. Due to the axisymmetry of the non-perturbed system, the equilibrium quantities do not depend on the azimuthal angle $\varphi$ and can in principle be pulled out of the $\varphi$-integral in expression \eqref{equ:deltaW} for $\delta W$. The displacement field on the contrary, is not constrained to be axisymmetric. In order to enable an easy analytical integration over $\varphi$ and reduce the integral to two dimensions, we Fourier-analyse every choice for $\vec \xi$ in terms of $\varphi$ with the mode index $m$.\par
For the purpose of testing the stabilising impact of so far neglected stellar properties, we start from displacement fields known to cause instabilities in a simpler model \citep{Tayler1, Tayler2}. \emph{Tayler instabilities} of the \emph{purely toroidal} field are present for
\begin{subequations}
\label{equ:XiTaylerToroidal}
\begin{align}
\xi_\varpi &=  X(\varpi,z) \e^{i m \varphi} \\
\xi_\varphi &=  \frac{iY(\varpi,z)}{m} \e^{im\varphi} \\
\xi_z &=  Z(\varpi,z) \e^{im\varphi},
\end{align}
\end{subequations}
depicted in cylindrical coordinates. \emph{Tayler instabilities} of the \emph{purely poloidal} field are present for
\begin{subequations}
\label{equ:XiTaylerPoloidal}
\begin{align}
\xi_\psi &=  \frac{X(\psi,\chi)}{\varpi B_\chi}  \e^{i m \varphi}\\
\xi_\varphi &=  \frac{i \varpi Y(\psi,\chi) }{m}\e^{i m \varphi} \\
\xi_z &=  B_\chi Z(\psi,\chi) \e^{i m \varphi},
\end{align}
\end{subequations}
depicted in toroidal coordinates. In both cases, $m \neq 0$ holds and $X,Y,Z$ are generally complex functions. Unlike \citet{Tayler1} and \citet{Tayler2}, we ensure $\delta W$ being real by taking the real part of the energy variational density in expression \eqref{equ:deltawi}, instead of taking the real part of $\vec \xi$ as defined in \eqref{equ:XiTaylerToroidal} and \eqref{equ:XiTaylerPoloidal}. This procedure allows for a consistent description comparable to the approaches of both \citet{Tayler1}, \citet{Tayler2} and \citet{Akguen}. Furthermore, we include the imaginary unit directly within definition \eqref{equ:XiTaylerPoloidal}, with the result of $\mathcal E$ being real for arbitrary choices of $X$ and $Y$. Both displacement fields describe an equivalent perturbation around the axis of the respective magnetic field: $\vec \xi$ as defined in cylindrical coordinates in \eqref{equ:XiTaylerToroidal} for the toroidal field case takes the same form as $\vec \xi$ expressed in toroidal coordinates, given in \eqref{equ:XiTaylerPoloidal}, for the 
poloidal field. 
Unless specified otherwise, the non-axisymmetric `kink' mode $m=1$ will be used, being believed to cause the strongest instability in a simple system with choice \eqref{equ:XiTaylerToroidal} for $\vec \xi$. A physical feature that is able to reduce the destabilising impact of this mode is a relevant candidate for explaining the stability of actual neutron stars.\par
Investigating stability in the case of \emph{mixed fields and stratification}, we will assume
\begin{subequations}
\label{equ:XiQuerAkguen}
\begin{align}
\bar \xi_r &= \hphantom{i}\tilde R(r,\vartheta) r \sin \vartheta \e^{i m\varphi}\\
\bar \xi_\vartheta &=\hphantom{i}\tilde S(r,\vartheta) r \sin \vartheta \e^{im\varphi}\\
\bar \xi_\varphi &= i\tilde T(r,\vartheta) r \sin \vartheta \e^{im\varphi},
\end{align}
\end{subequations}
depicted in spherical coordinates, as used in \citet{Akguen} in order to make our result comparable to this work. $\tilde S$, $\tilde R$, $\tilde T$ are generally complex dimensionless functions. Assumptions \eqref{equ:XiTaylerToroidal} and \eqref{equ:XiQuerAkguen} imply the same $\varphi$-dependence and are structurally equivalent if the axisymmetric generating functions are related via
\begin{subequations}
\label{equ:XYZRSTRelation}
\begin{align}
 X &= \left(\varpi \tilde R + z \tilde S \right) \frac{\varpi}{r}\\ 
 Y &= m \varpi \tilde T\\
 Z &= \left(z \tilde R - \varpi \tilde S \right) \frac{\varpi}{r}.
\end{align}
\end{subequations}
However, in order to show the stabilising impact of the poloidal magnetic field component on the toroidal one, \citet{Akguen} focus their investigations on a spatially constrained area $A$ where the poloidal field provides a positive contribution to the energy variation whereas the toroidal field is unstable. This concept is implemented by a localised displacement field, vanishing everywhere outside $A$:
\begin{equation}
\label{equ:XiAkguen}
 \vec \xi = \begin{cases}
             \vec{\bar \xi} \quad &\text{for} \quad \left[\left(\frac{r-r_0}{\delta_r}\right)^2 + \left( \frac{\vartheta-\vartheta_0}{\delta_\vartheta}\right)^2 \right] < 1\\
             0 \quad &\text{else.}
            \end{cases}
\end{equation}
$(r_0,\vartheta_0)$ and $\delta_r$, $\delta_\vartheta$ determine position and spatial extend of $A$.\par
Note that choices \eqref{equ:XiTaylerToroidal} and \eqref{equ:XiAkguen} for $\vec \xi$ describe a perturbation around the stellar axis, whereas choice \eqref{equ:XiTaylerPoloidal} is a perturbation around the magnetic axis, cf. Fig. \ref{fig:TorPol}. The defining functions $X$, $Y$, $Z$ and $\tilde R$, $\tilde S$, $\tilde T$ can be assumed to be real without restriction. Their real and imaginary parts contribute nothing but equal terms to the energy integral \citep{Tayler1} as we show in section \ref{Applications}. Thus, a potential imaginary part wouldn't change the structure or sign of the integral.

\subsection{Cowling approximation}
\label{CowlingApproximation}
Analytical stability studies \citep{Tayler1, Tayler2, Tayler3, Tayler4, Akguen} on compact stars commonly assume a negligible Eulerian perturbation of the gravitational potential, shown in \eqref{equ:deltaPhi}, in order to simplify the energy variation \eqref{equ:deltaW} by removing $\delta_{\text{grav}}^{\text{nC}}$. The gravitational contribution to the energy variational density simplifies to $\mathcal E_{\text{grav}} \approx \mathcal E^{\text{Cowl}}_{\text{grav}}$ if Cowling approximation,
\begin{equation}
\label{equ:CowlingApproximation}
\delta \Phi = 0 \quad \Rightarrow \quad \mathcal E^{\text{nC}}_{\text{grav}} = 0,
\end{equation}
is used. That way, additional integrations being necessary to calculate $\delta \Phi$ can be avoided. Nevertheless, it is not obviously clear for every displacement field that the application of Cowling's approximation doesn't influence the resulting stability behaviour. For a kink mode type displacement around the stellar axis, as in \eqref{equ:XiTaylerToroidal} and \eqref{equ:XiAkguen}, Cowling approximation tends to let the system appear more stable than it actually is \citep{Tayler1}, being a valid simplification for proving instabilities in a simple model. On the contrary, showing the stabilising impact of realistic stellar properties in Cowling approximation can potentially be distorted by neglecting $\delta W^{\text{nC}}_{\text{grav}} <0$ if its absolute value is of the order of the total energy variation in Cowling approximation: $|\delta W^{\text{nC}}_{\text{grav}}| \sim |\delta W |$.\par
If Cowling's approximation is dropped, $\delta \Phi$ given by \eqref{equ:deltaPhi} has to be expressed explicitly. This has been done by \citet{Chandrasekhar} and \citet{ChandrasekharLebovitz} in the case of a spherically symmetric system with $\vec \xi$ being expressed by eigenfunctions. We follow this approach, deriving $\delta \Phi$ in an axisymmetric system with choice \eqref{equ:XiQuerAkguen} for the displacement field. 
Making use of the Tayler expansion for $\frac{1}{|\vec r - \vec r'|}$, the Eulerian perturbation of the gravitational potential can be written as
\begin{equation}
\label{equ:deltaPhiAnsatz}
\delta \Phi(\vec r) = \sum\limits_{l=0}^\infty \delta \Phi_l(r) Y_l^m(\vartheta, \varphi)
\end{equation}
with $Y_l^m(\vartheta,\varphi)$ being spherical harmonics and
\begin{equation}
\label{equ:deltaPhil}
 \delta \Phi_l(r) = 2 \pi \frac{(l-|m|)!}{(l+|m|)!} \left\{ \frac{J_l(r)}{r^{l+1}} - r^l K_l(r) \right\}.
\end{equation}
For usage of expression \eqref{equ:deltaPhiAnsatz} in $\mathcal E^{\text{nC}}_{\text{grav}}$, given in \eqref{equ:deltawgravnC}, we take into account first and second order terms in the expansion. With our choice of $m=1$, the zeroth order vanishes, such that
\begin{equation}
\label{equ:deltaPhiErgebnisSpez}
\delta \Phi(\vec r) = \delta \Phi_1(r) Y_1^1(\vartheta, \varphi) + \delta \Phi_2(r) Y_2^1 (\vartheta,\varphi) + \mathcal O \left(\frac{r_<^3}{r_>^{4}} \right),
\end{equation}
with $r_< = \text{min}(r,r')$ and $r_> = \text{max}(r,r')$. $J_l(r)$ and $K_l(r)$ are two-dimensional integrals given by the $\vec \xi$-defining functions $\tilde R$, $\tilde S$, $\tilde T$ defined in \eqref{equ:XiQuerAkguen}:
\begin{subequations}
\label{equ:JlKl2D}
\begin{align}
 J_l(r) &= \iint\limits_{S_J(r)} \rho_0(r') r'^{l-1} \left[ \left(l\tilde R(r',\vartheta') + \frac{m\tilde T(r', \vartheta')}{\sin\vartheta'} \right) Y_l^m(real)(\vartheta') \right.\\
  &\nonumber \hphantom{=\iint\limits_0^r}+\tilde S(r',\vartheta') \partial_{\vartheta'} Y_l^m(real) (\vartheta') \bigg] r'^3 \sin^2 \vartheta' \d \vartheta' \d r'\\
K_l(r) &= \iint\limits_{S_K(r)} \frac{\rho_0(r')}{r'^{l+2}} \left[ \left((l+1)\tilde R(r',\vartheta') - \frac{m\tilde T(r', \vartheta')}{\sin\vartheta'} \right) Y_l^m(real)(\vartheta') \right.\\
&\nonumber \hphantom{=\iint\limits_0^r}-\tilde S(r',\vartheta') \partial_{\vartheta'} Y_l^m(real) (\vartheta') \bigg] r'^3 \sin^2 \vartheta' \d \vartheta' \d r'.
\end{align}
\end{subequations}
The integration areas are subdomains of an area enclosed by the semicircle at $z>0$ with radius $R$. In the case of $S_J(r)$, $r'$ runs from $0$ to $r$; in the case of $S_K(r)$, $r'$ runs from $r$ to $R$. The integration will be carried out in spherical coordinates.

\section{Numerical procedure}
\label{NumericalProcedure}
For the calculation of the energy variation defined in \eqref{equ:deltaW}, the corresponding density given in \eqref{equ:deltawi} has to be integrated over the investigated area, for instance the stellar volume or a potentially unstable region. The integrand contains position variables, equilibrium quantities $Q_0$ and their spatial derivatives, the displacement field $\vec \xi$ and in full non-Cowling treatment $J_l$ and $K_l$.\par
First, we calculate $m_0(r)$, $p_0(r)$, $\rho_0(r)$, $\Phi_0(r)$ by solving the system equations \eqref{equ:SystemEquations} with the background equation of state \eqref{equ:EquationOfStateBackground} using a classical Runge-Kutta method. They are interpolated to $m_0(\varpi,z)$, $p_0(\varpi,z)$, $\rho_0(\varpi,z)$, $\Phi_0(\varpi,z)$ making use of cubic splines. Next, magnetic field $\vec B$ and displacement field $\vec \xi$ are parametrized as defined in sections \ref{MagneticField} and \ref{DisplacementField}. $J_l$ and $K_l$ follow from integrating expression \eqref{equ:JlKl2D} with a two-dimensional Simpson's method. Finally, $\mathcal E$ can be integrated. The angular integration over $\varphi$ is carried out analytically, the remaining integration numerically using Simpson's method in two dimensions. For a detailed description of the numerical procedure see appendix \ref{NumericalProcedureInDetail}.

\section{Applications}
\label{Applications}
\subsection{Purely toroidal magnetic field}
\label{PurelyToroidalMagneticField}
In a first test we apply the semi-analytic stability analysis scheme on a simple neutron star model with a \emph{purely toroidal magnetic field}. The equilibrium state of this system is known to be unstable. We use the model system defined in section \ref{ModelSystem} without stratification, i.e. $\Gamma_1 = \Gamma_0$, and in Cowling approximation, i.e. $\mathcal E_{\text{grav}}^{\text{nC}} = 0$ as described in section \ref{CowlingApproximation}. Magnetic field and displacement field are chosen according to assumptions \eqref{equ:BTaylerToroidal} and \eqref{equ:XiTaylerToroidal}. The integration area is defined by expression \eqref{equ:IntegrationAreaV} in the appendix.

\subsubsection{Setup}
\label{PurelyToroidalMagneticFieldSetup}
Inserting the above choice for $\vec \xi$ into the general form of $\delta W$, given in \eqref{equ:deltaW} with \eqref{equ:deltawi}, the explicit energy variation that has to be studied in the purely toroidal field case follows:
\begin{equation}
\label{equ:deltaWexplicitBtor}
 \delta W = \frac{\pi}{2} I(X_R, Z_R) + \frac{\pi}{2} I(X_I, Z_I)
\end{equation}
with
\begin{align}
\label{equ:deltaWexplicitBtorB}
 I(\mathcal X, \mathcal Z) &= \displaystyle\iint\limits_{S(\text{star})} \left\{ B_\varphi^2 \left(\varpi \partial_\varpi \left( \frac{\mathcal{X}}{\varpi}\right) + \partial_z \mathcal{Z} \right)^2 + B_\varphi^2 \frac{\mathcal{Z}^2 - \mathcal{X}^2}{\varpi^2}  \right.\\
 \nonumber &\hphantom{\displaystyle\iint\limits_{S(\text{star})}}  -2 \mathcal{X}  B_\varphi \cdot \frac{\mathcal{X}\partial_\varpi B_\varphi + \mathcal{Z}\partial_z B_\varphi}{\varpi}   + \left(g_\varpi \mathcal{X} + g_z \mathcal{Z} \right)\\
 \nonumber &\hphantom{\displaystyle\iint\limits_{S(\text{star})}} \cdot  \left[ -\frac{\rho_0}{\Gamma_0 p_0} \left( B_\varphi \left( \frac{\mathcal{X} \partial_\varpi (\varpi B_\varphi)}{\varpi} + \mathcal{Z} \partial_z B_\varphi \right) \right.\right.\\
 \nonumber &\hphantom{\displaystyle\iint\limits_{S(\text{star})}} \left.  + \rho_0 \left(g_\varpi \mathcal{X}+ g_z \mathcal{Z} \right) \Bigg) +  \left( \mathcal{X} \partial_\varpi \rho_0 + \mathcal{Z} \partial_z \rho_0 \right) \Bigg] \vphantom{\left(\frac{\mathcal{X}}{\varpi}\right)^2 } \right\} \varpi \d \varpi \d z,
\end{align}
where $g_\varpi = \pm \partial_\varpi \Phi_0$, $g_z = \pm \partial_z \Phi_0$, defined in \eqref{equ:GravitationalField}. Real and imaginary parts of the $\vec \xi$-defining functions are denoted by the indices $R$ and $I$. The integration area $S(\text{star})$ is one half of the stellar cross section in the $(\varpi, z)$-plane for constant $\varphi$. This area corresponds to the integration surface remaining after the $\varphi$-part of the 3D-integration extending over the stellar volume has been carried out. It is defined in expression \eqref{equ:IntegrationAreaV} of the appendix. The integrand $\mathcal E$ of the energy variation has been minimized with respect to $\mathcal Y$ in order to achieve \eqref{equ:deltaWexplicitBtorB}, as shown in \citet{Tayler1}. The minimising value
\begin{equation}
\label{equ:Ymin}
 \mathcal Y_{\text{min}} = \partial_\varpi \left(\varpi \mathcal X\right) + \varpi \partial_z \mathcal Z + \frac{\varpi \rho_0}{\Gamma_0 p_0} \left(g_\varpi \mathcal X + g_z \mathcal Z \right)
\end{equation}
yielding $\mathcal E_{\text{min}} = \mathcal E(\mathcal Y_{\text{min}})$ has been inserted.\par
One can see from expression \eqref{equ:deltaWexplicitBtor} that real and imaginary parts of the $\vec \xi$-defining functions contribute equal parts to $\delta W$. This holds for the other cases of purely poloidal field and mixed field with stratification as well. Thus, we will assume $X=X_R$, $Y=Y_R$, $Z=Z_R$ and $\tilde R=\tilde R_R$, $\tilde S=\tilde S_R$, $\tilde T=\tilde T_R$ to be real hereafter. Furthermore, in this section we assume a simple form
\begin{equation}
\label{equ:ChoiceTaylerToroidalXZ}
 X =1 \quad Z=0,
\end{equation}
letting the displacement field lie in the $(\varpi,\varphi)$-plane. Alternatively, arbitrary parametrized functions of $\varpi$ and $z$ can be used which we abstain from here. Actually, we don't expect a change in the stability behaviour for different choices of $X$ and $Z$ as \citet{Tayler1} showed instability for arbitrary assumptions. Nevertheless, the parametrisation of $\vec \xi$ is a promising tool for future investigations of possibly unstable modes.\par
Splitting equation \eqref{equ:deltaWexplicitBtorB} into separate parts stemming from magnetic field, fluid pressure and gravity, the summands of the energy variation read
\begin{subequations}
\label{equ:deltaWiexplicitBtor}
\begin{align}
 \delta W_{\text{magn}} &= \frac{\pi}{2} \displaystyle\iint\limits_{S(\text{star})} \left[\frac{X^2 B_{\varphi}^2}{\varpi^2}+\left(\partial_\varpi(XB_{\varphi})\right)^2-\frac{X}{\varpi} \partial_\varpi (\varpi B_{\varphi}) \right.\\
 &\nonumber \hphantom{= \frac{\pi}{2} \displaystyle\iint} \left. \cdot\partial_\varpi (XB_{\varphi}) - \frac{XY_{\text{min}}B_{\varphi}}{\varpi^2} \partial_\varpi (\varpi B_{\varphi}) \right] \varpi \d \varpi \d z\\
\label{equ:deltaWfluidexplicitBtor}
 \delta W_{\text{fluid}} &= \frac{\pi}{2} \displaystyle\iint\limits_{S(\text{star})} \left[ \Gamma_0 p_0 \left(\frac{\partial_\varpi(\varpi X)-Y_{\text{min}}}{\varpi} \right)^2 \right.\\
 &\nonumber \hphantom{= \frac{\pi}{2} \displaystyle\iint} \left. + X \partial_\varpi p_0 \frac{\partial_\varpi(\varpi X)-Y_{\text{min}}}{\varpi} \right] \varpi \d \varpi \d z\\
\label{equ:deltaWiexplicitBtorGrav}
\delta W_{\text{grav}}^{\text{Cowl}} &= \frac{\pi}{2} \displaystyle\iint\limits_{S(\text{star})} \left[ X^2 g_\varpi \partial_\varpi \rho_0 + X \rho_0 g_\varpi \frac{\partial_\varpi(\varpi X)-Y_{\text{min}}}{\varpi}  \right] \varpi \d \varpi \d z.
\end{align}
\end{subequations}
Contrary to the assumption of \citet{Tayler1}, we neglected the magnetic field contribution in the hydrostatic equilibrium equation \eqref{equ:HydrostaticEquilibrium} for rearranging the terms in the integrand. Taking it into account would cause an inconsistency as we already chose to treat the equilibrium system as an unmagnetised background.\par
Before integrating, we solve the system equations \eqref{equ:SystemEquations} creating different neutron star models for a polytropic index of $\Gamma_0=2$, presented in Table \ref{tab:TOVparameters}. Here and in all following applications we work in a Newtonian framework with the following justification. Whereas the negligence of relativistic terms changes the absolute value of the energy variation $\left|\delta W \right|$ indeed, we are interested in critical parameter values, for which $\delta W (\text{parameters}) \gtrless 0$ changes sign. The location of these critical points is typically less influenced by relativistic corrections than the absolute value is. 
\begin{table}
	\centering
	\caption{Parameter values for central density $\rho_c$ and proportionality constant $\kappa$ with corresponding stellar masses $M$ and radii $R$, being used in section \ref{PurelyToroidalMagneticField}. $M$ and $R$ have been constructed solving the system equations \eqref{equ:SystemEquations} with $\Gamma_0=2$.}
	\label{tab:TOVparameters}
	\begin{tabular}{llll} 
		\hline
		$\rho_c$ in \SI{e15}{g cm^{-3}} &$ \kappa$ in \SI{e11}{cm^2}  &$M$ in $\text{M}_\odot$ &$R$ in \si{km}\\
		\hline
		\num{1.0} &\num{8} &\num{0.90} &\num{11.2} \\
		\num{1.2} &\num{8} &\num{1.08} &\num{11.2} \\
		\num{1.0} &\num{10} &\num{1.26} &\num{12.5} \\
		\num{1.1} &\num{10} &\num{1.39} &\num{12.5} \\
		\hline
	\end{tabular}
\end{table}

\subsubsection{Results}
\begin{figure}
	\includegraphics[width=\columnwidth]{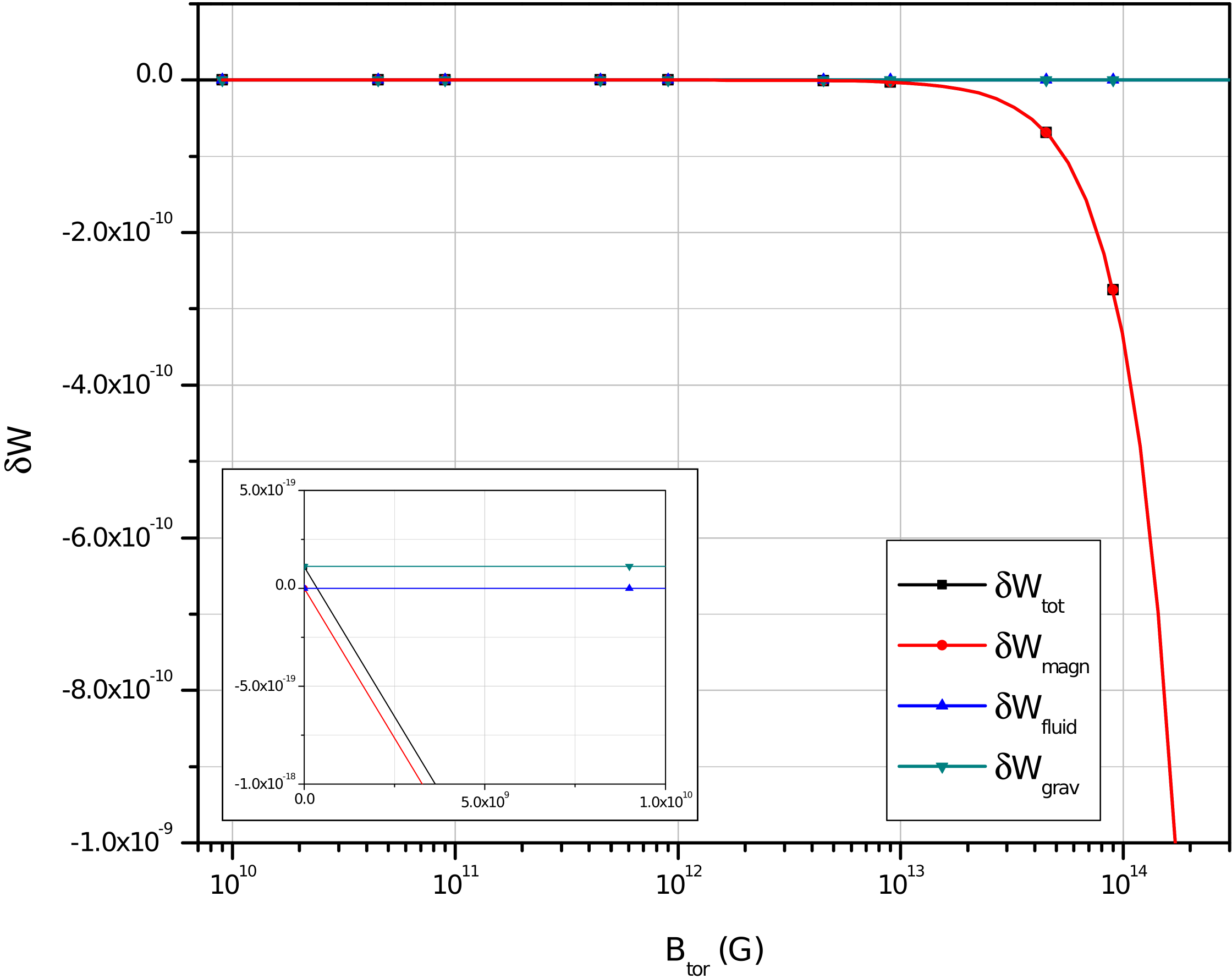}
    \caption{Calculated energy variation $\delta W$ in a simple system with a purely toroidal magnetic field for $M=\num{1.26}$\,$\text{M}_\odot$, $R=\SI{12.5}{km}$. The plot shows the total energy variation and its constituents stemming from magnetic field, fluid pressure and gravity. $\delta W$ is positive for $B_\text{tor} =0$ and negative for all other field strengths. For the dependency of the energy variation on the magnetic field strength, see Fig.~\ref{fig:deltaWresultBtor}.}
\label{fig:deltaWsignBtor}
\end{figure}

\begin{figure}
	\includegraphics[width=\columnwidth]{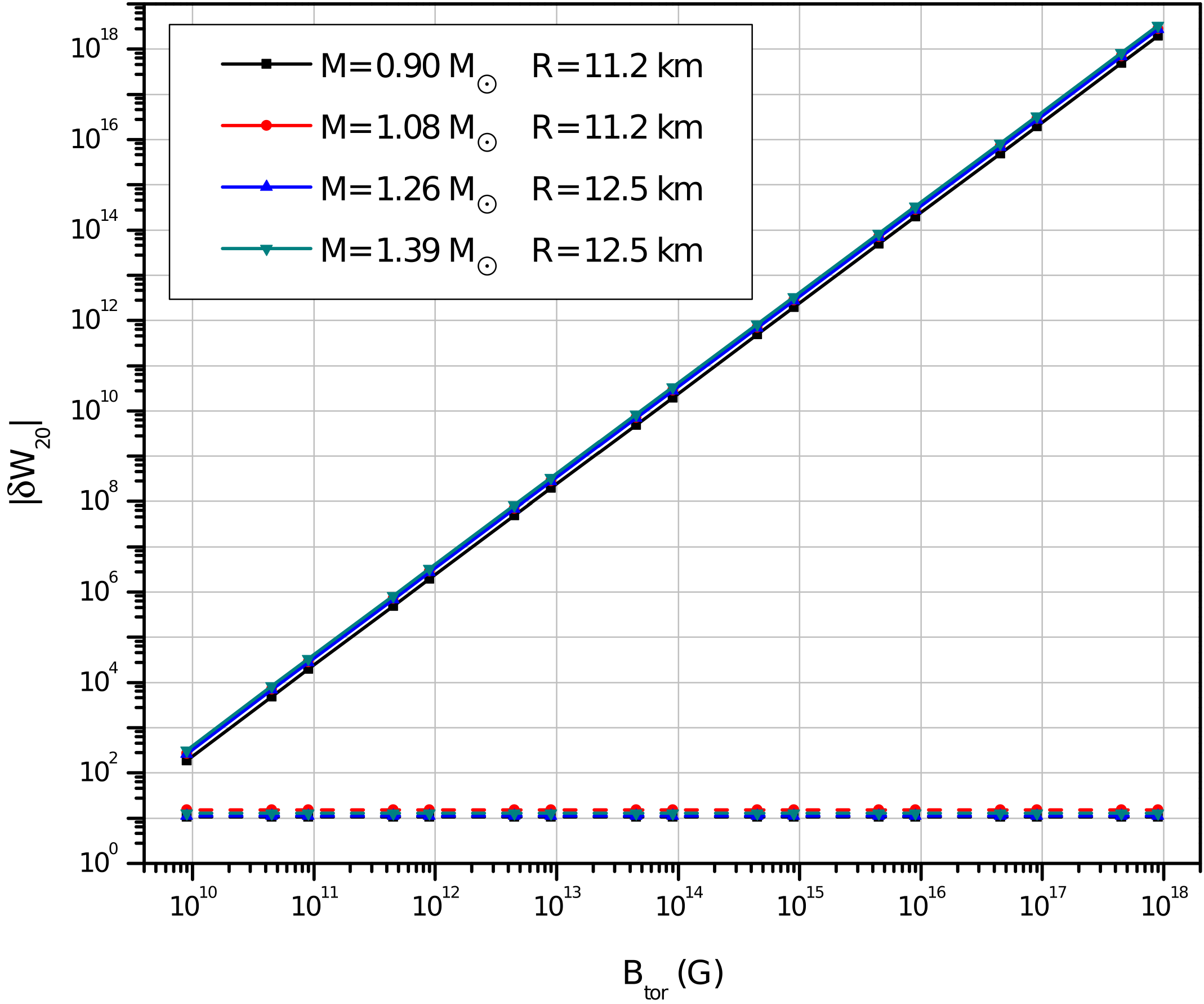}
    \caption{Behaviour of the absolute value of the energy variation in a simple system with a purely toroidal magnetic field. The plot shows $\left| \delta W_{20}\right| \equiv \num{e20} \left|\delta W\right|$ for the total energy variation $\delta W$ (solid line) and its gravitational part $\delta W_\text{grav}$ (dashed line). Note that $\delta W<0$, $\delta W_{\text{magn}}<0$, $\delta W_{\text{fluid}} = 0$, $\delta W_{\text{grav}}>0$ $\forall$ $B_\text{tor} \neq 0$, cf. Fig.~\ref{fig:deltaWsignBtor}. The behaviour of the total energy variation is driven by the magnetic field contribution. For $B_\text{tor} =0$, $\delta W$ lies between $\num{1.1}$ and $\num{1.5e-19}$ for the models under consideration. All parameters being used are defined in section \ref{PurelyToroidalMagneticField}.}
\label{fig:deltaWresultBtor}
\end{figure}

$\delta W$ as given by \eqref{equ:deltaWexplicitBtor}, \eqref{equ:deltaWexplicitBtorB} as well as $\delta W_{\text{magn}}$, $\delta W_{\text{fluid}}$ and $\delta W_{\text{grav}}^{\text{Cowl}}$ as given by \eqref{equ:deltaWiexplicitBtorGrav} have been calculated for varying the toroidal magnetic field strength $B_{\text{tor}}$ from \SIrange[range-units=single]{e10}{e18}{G} and $B_{\text{tor}}=0$. Fig.~\ref{fig:deltaWsignBtor} illustrates the sign of the energy variation. The behaviour of the total energy variation with $B_\text{tor}$ is shown in Fig.~\ref{fig:deltaWresultBtor}, where the individual terms are discussed as well.\par
All considered models in Fig.~\ref{fig:deltaWresultBtor} show the same stability behaviour: The total energy variation is slightly positive for $B_{\text{tor}}=0$, implying a stable or marginally stable unmagnetised system. The small absolute values of the energy variation we calculated for $B_\text{tor}=0$ most probably lie within the error range of the numerical integration. For every nonzero magnetic field, the energy variation is negative independently of the field strength, as can be seen from Fig.~\ref{fig:deltaWsignBtor}. Thus, the semi-analytic results indicate that $\delta W <0$ $\forall$ $B_\text{tor}$ and $\delta W \to 0$ for $B_\text{tor} \to 0$. Therefore, the critical toroidal field strength above which the energy variation is negative is $B_\text{tor}^\text{crit}=0$. This behaviour verifies the result of an instability which is solely driven by the magnetic field geometry and has been found by \citet{Tayler1}.\par
$\delta W$ decreases with increasing magnetic field strength, caused by the stronger negative contribution from $\delta W_{\text{magn}}$, see Fig.~\ref{fig:deltaWsignBtor}. The fluid pressure contribution $\delta W_{\text{fluid}}$ vanishes for all field strengths. This can be clarified as follows. Using hydrostatic equilibrium equation \eqref{equ:HydrostaticEquilibrium} and the minimising value for $Y$ chosen in \eqref{equ:Ymin}, the relation
\begin{equation}
 X \partial_\varpi p_0  = \rho_0 X g_\varpi  = - \Gamma_0 p_0 \frac{\partial_\varpi (\varpi X) - Y_{\text{min}}}{\varpi} 
\end{equation}
holds if $Z=0$. Thus, from the explicit form of $\delta W_{\text{fluid}}$ in \eqref{equ:deltaWfluidexplicitBtor}, $\delta W_{\text{fluid}} = 0$ follows. The gravitational contribution in Fig.~\ref{fig:deltaWsignBtor} and Fig.~\ref{fig:deltaWresultBtor} is slightly positive and constant for all field strengths. Therefore, the total hydrostatic contribution is small and positive, implying a stable system if the star is not magnetised. The instability is indeed caused by the toroidal magnetic field independently of its strength, verifying the statement of \citet{Tayler1}.

\subsection{Purely poloidal magnetic field}
\label{PurelyPoloidalMagneticField}
For the purpose of investigating a simple system with a \emph{purely poloidal magnetic field}, we use the model system described in section \ref{ModelSystem}. We neglect stratification and assume Cowling approximation, as already done in section \ref{PurelyToroidalMagneticField}. Magnetic field and displacement field are chosen according to assumptions \eqref{equ:BTaylerPoloidal} and \eqref{equ:XiTaylerPoloidal}. The integration area is defined by expression \eqref{equ:IntegrationAreaTorus} in the appendix.

\subsubsection{Setup}
In accordance with \citet{Tayler2}, we choose the $\vec \xi$-defining functions as
\begin{subequations}
\label{equ:ChoiceTaylerPoloidalXYZ}
\begin{align}
 X &= X_0 \sin \chi\\
 Y &= \frac{X_0 \sin \chi}{\varpi R_{\text{tor}}} + \frac{X_0 \sin \chi}{r_{\text{tor}}^2}\\
 Z &= X_0 \sin \chi \frac{r_{\text{tor}}-R_{\text{tor}} \cos \chi}{r_{\text{tor}}^2 R_{\text{tor}}}
\end{align}
\end{subequations}
with $X_0=1$. That way,
\begin{subequations}
 \label{equ:XiBpolAnnahmen}
 \begin{align}
\vec \nabla \cdot \vec \xi &= 0\\
\vec \xi \cdot \vec \nabla \Phi_0 &= 0,
\end{align}
\end{subequations}
what causes all terms besides the magnetic contribution in $\delta W$ to vanish, keeping in mind that we neglect $\mathcal E_{\text{grav}}^{\text{nC}}$, cf. expression \eqref{equ:deltawi}. Subsequently, the simplified energy variation is
\begin{align}
\label{equ:deltaWexplicitBpol}
\delta W = \delta W_{\text{\text{magn}}} &= \frac{\pi}{2} \displaystyle\iint\limits_{S(\text{torus})} \left[ \frac{(\partial_\chi X)^2}{r_{\text{tor}} \varpi} + \frac{\varpi}{m^2} (\partial_\chi Y )^2 + \frac{r_{\text{tor}}^2}{\varpi} Y^2 \right. \\
&\nonumber \hphantom{\frac{\pi}{2}\displaystyle\iint\limits_{S(\text{torus})}} \left. + \frac{r_{\text{tor}} \cos \chi + 2 \varpi}{\varpi^2} \left( \frac{Z \partial_\chi X}{\varpi} - XY\right) \right] \d \varpi \d z.
\end{align}
The integration area $S(\text{torus})$ is the cross section of a torus in the $(\varpi, z)$-plane for constant $\varphi$. It is defined in expression \eqref{equ:IntegrationAreaTorus} of the appendix. The torus lies within the star, including the region of poloidal field lines closing inside the star, cf. Fig.~\ref{fig:ToroidalCoordinates} and Fig.~\ref{fig:TorPol}. The original 3D-integration area before integrating over $\varphi$ extended over the volume of this torus.\par
During the calculation, we vary the mode index $m$. The parameters used for computation are $\rho_c =$ \SI{e15}{g.cm^{-3}}, $\Gamma_0=2$ and $\kappa=$ \SI{100}{cm^2}, eventuating in a neutron star model of $M=\num{1.26}$\,$\text{ M}_{\odot}$, $R=\SI{12.25}{km}$ as calculated by solving the Newtonian system equations \eqref{equ:SystemEquations}. We determine the centre of the integration area by $R_{\text{tor}}=R/2$. We also vary the size of the integration area from $r_{\text{tor}}/R_{\text{tor}}=$ \num{0.6} to \num{0.9634}. The upper limit corresponds to the maximum value for which the energy variation is still negative and the instability still detectable. Thus, in all cases the toroidal integration area fully lies within the star.

\subsubsection{Results}
\begin{figure}
	\includegraphics[width=\columnwidth]{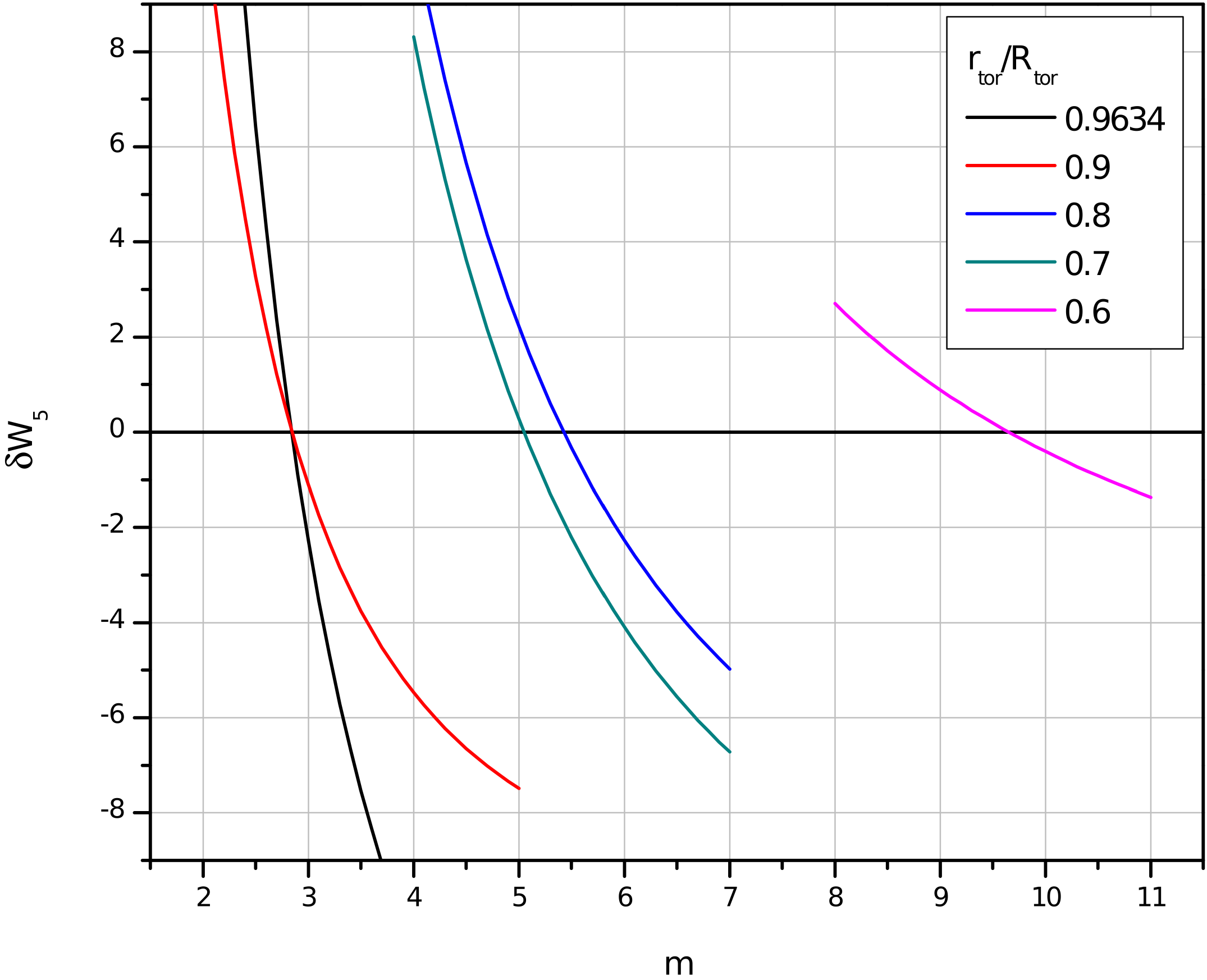}
    \caption{Calculated total energy variation in a simple system with a purely poloidal magnetic field. $\delta W_5 \equiv \num{e5}\delta W$ is plotted against the mode index $m$ of the displacement field for different sizes $r_\text{tor}/R_\text{tor}$ of the integration area. All parameters being used are defined in section \ref{PurelyPoloidalMagneticField}.}
\label{fig:deltaWresultBpol}
\end{figure}

The calculated total energy variation in the poloidal field case is plotted as a function of the mode index $m$ in Fig.~\ref{fig:deltaWresultBpol} for different sizes of the integration area, specified by $r_{\text{tor}}/R_{\text{tor}}$. In all cases, the energy variation is positive for small $m$ and becomes negative with increasing $m$. That means, one can always find a mode $m$ for which the system is unstable against $\vec \xi$, as given by \eqref{equ:XiTaylerPoloidal}. This result implies the generic instability of purely poloidal fields in neutron stars found by \citet{Tayler2}. Also the behaviour of $\delta W(m)$ is in accordance with their result since $m \rightarrow \infty$ has been assumed in the instability proof there. The instability is a structural one, not depending on the field strength $B_{\text{pol}}$, as can be inferred from the fact that $B_{\text{pol}}$ cancelled out of the energy variation \eqref{equ:deltaWexplicitBpol}.\par
The different integration areas give different critical mode indices for which $\delta W$ changes sign. This is caused by the fact that different amounts of spatial areas contributing positive and negative energy variational densities are included in the integral. Keep in mind that the energy variational density has been integrated over a torus shaped volume for all integration areas. These tori are centred around the magnetic axis and only differ in their radial extends $r_{tor}$. All choices for $r_\text{tor}$ exclude the region of diverging positive contributions close to the symmetry axis as requested. Close to the magnetic axis, areas providing positive and areas providing negative contributions with large absolute values exist. Their spatial extend varies with $m$. For big mode indices, such as $m=11$ for instance, these regions fully lie within all of the assumed integration areas. For small $m$, smaller integration areas with $r_\text{tor}/R_\text{tor} \lesssim \num{0.4}$ may not contain the whole region but cut the outside edges of the area with large absolute values. Therefore, the critical mode index depends on the size of the integration area in a non-trivial way. This can be seen by plotting the integrand depending on the spatial variables. The qualitative behaviour in Fig.~\ref{fig:deltaWresultBpol}, however, is equivalent in all cases. That means, the outcome of the semi-analytic stability investigation is independent of the chosen integration area. This holds as long as the integration area includes the possible instability region and excludes areas with open field lines and strong positive $\mathcal E$-contributions that would overlay the negative contribution caused by the actual instability.\par
One consequence of this result is: At least for perturbation modes of the type fulfilling assumptions \eqref{equ:XiBpolAnnahmen}, simple neutron star models with purely poloidal fields host instabilities that cannot be stabilised by fluid pressure or gravity contributions for any field strength. Note that the stability analysis we provide here is based on a stationary picture. Even if higher mode indices result in more negative energy variations, this doesn't say anything about the fastest growing mode, often referred to as the `strongest instability'.

\subsection{Mixed field and stratification}
\label{MixedFieldAndStratification}
Investigating \emph{mixed fields and stratification}, we follow the approach of \citet{Akguen} in order to achieve comparable results. The system defined in section \ref{ModelSystem} is used. Note that the form of $\mathcal E_{\text{fluid}} + \mathcal E_{\text{grav}}^{\text{Cowl}}$, defined in \eqref{equ:deltawfluid} and \eqref{equ:deltawgravCowl}, is equivalent to the hydrostatic energy variational density in \citet{Akguen}, as shown in the appendix \ref{EquivalenceWithAkguen}. Magnetic field and displacement field are chosen according to assumptions \eqref{equ:BAkguen} and \eqref{equ:XiAkguen} with \eqref{equ:XiQuerAkguen}. The integration area is defined in expression \eqref{equ:IntegrationAreaV}.\par
As mentioned in section \ref{CowlingApproximation}, for displacement fields of the type assumed here, a system in Cowling approximation appears more stable than it actually is. \citet{Akguen} show how the Tayler instability of the toroidal field can be stabilised in the presence of stratification and a poloidal field with a minimum field strength using Cowling approximation. Addressing the question whether this result is strongly influenced or even suppressed in full treatment, we consider the problem both with and without Cowling approximation.

\subsubsection{Setup}
In analogy to section \ref{PurelyToroidalMagneticFieldSetup}, the energy variational density has been minimized with respect to the $\varphi$-component of the displacement field, yielding
\begin{equation}
\tilde T_{\text{min}} = -\frac{E_1}{2 m E_2}
\end{equation}
with
\begin{align}
 E_1 &= -2 \varpi \Gamma_1 p_0 D_0 - \varpi\tilde R \left(\partial_r p_0 + \rho_0 \partial_r \Phi \right) - \frac{\beta \Lambda(\beta)}{4 \pi \varpi},\\
 E_2 &= \Gamma_1 p_0
\end{align}
and
\begin{equation}
\Lambda(U)=\tilde R \partial_r U + \frac{\tilde S \partial_\vartheta U}{r}  \qquad D_0 = \frac{3 \tilde R}{r} + \partial_r \tilde R + \frac{2z \tilde S}{\varpi r} + \frac{\tilde S}{r},
\end{equation}
following the notation of \citet{Akguen}. The remaining $\vec \xi$-defining functions are chosen as
\begin{subequations}
 \label{equ:ChoiceAkguenRS}
\begin{equation}
\tilde R = \begin{cases}  - \frac{\xi_0}{r} \sigma \left(1-\bar\chi^2\right)^{\sigma-1} \partial_\vartheta \bar\chi^2 \quad &\text{for} \quad \bar \chi^2 < 1\\
             0 \quad &\text{else}
             \end{cases}
\end{equation}
and
\begin{equation}
\tilde S =\begin{cases}
	     \frac{\xi_0}{R} \sigma \left(1-\bar\chi^2\right)^{\sigma-1} \partial_x \bar\chi^2 
           &\text{for} \quad \bar \chi^2 < 1\\
             0 \quad &\text{else,}
          \end{cases}
\end{equation}
\end{subequations}
where $\bar \chi^2 = \left[\left(\frac{r-r_0}{\delta_r}\right)^2 + \left( \frac{\vartheta-\vartheta_0}{\delta_\vartheta}\right)^2 \right]$. We use the parameter values chosen in \citet{Akguen}: $\xi_0 =1$, $\sigma=\frac{3+\sqrt{3}}{2}$. The explicit form of the energy variation can be written as
\begin{equation}
\label{equ:deltaWexplicitAkguen}
 \delta W = \frac{\pi}{2} \displaystyle\iint\limits_{S(\text{star})} \left[ E_0 - \frac{E_1^2}{4 E_2} + 2 E_{\text{pol}} \right] \varpi \d \varpi \d z
\end{equation}
with
\begin{align}
 E_0 &= \left[\varpi^2 \Gamma_1 p_0 + \frac{\beta^2}{4\pi} \right] D_0^2 \\
 \nonumber &+ \left[ \varpi^2 \tilde R \left(\partial_r p_0 + \rho_0 \partial_r \Phi \right) + \frac{\beta \Lambda(\beta)}{4 \pi} -\beta^2 \frac{\varpi \tilde R + z \tilde S}{\pi \varpi r} \right] D_0\\
 \nonumber &+ \left[\varpi^2 \tilde R^2 \partial_r \rho  \partial_r \Phi - \beta \Lambda(\beta) \frac{\varpi \tilde R + z \tilde S}{2 \pi \varpi r} + \beta^2 \left(\frac{\varpi \tilde R+z \tilde S}{\varpi r} \right)^2  \right.\\
 \nonumber &\qquad\hphantom{+\left[\varpi^2 \tilde R^2 \partial_r \rho  \partial_r \Phi - \beta \Lambda(\beta) \frac{\varpi \tilde R + z \tilde S}{2 \pi \varpi r} \right.}  \left.+ m^2 \beta^2 \frac{\tilde R^2+ \tilde S^2}{2 \pi \varpi^2} \right],\\
 E_{\text{pol}} &= \frac{1}{8 \pi \varpi^2 r^2} \left[ r^2 \left(m \tilde T \partial_r \alpha - \frac{\varpi \Lambda (\alpha)}{r} - \varpi \partial_r \Lambda(\alpha) \right)^2 \right.\\
 \nonumber & + \varpi r^2 \left( m \tilde T \partial_r \alpha - \frac{\varpi \Lambda(\alpha)}{r} - \varpi \partial_r \Lambda(\alpha) \right)  \tilde R \Delta^* \alpha\\
 \nonumber & + \left( m \tilde T \partial_\vartheta \alpha - z \Lambda (\alpha) - \varpi \partial_\vartheta \Lambda(\alpha) \right)^2 + \varpi r \left(  m \tilde T \partial_\vartheta \alpha - z \Lambda(\alpha) \right.\\
 \nonumber & - \varpi \partial_\vartheta \Lambda(\alpha) \Big) \tilde S \Delta^* \alpha + \varpi^2 \left( \partial_r \tilde T \partial_\vartheta \alpha - \partial_\vartheta \tilde T \partial_r \alpha \right)^2 \Bigg]
\end{align}
and
\begin{equation}
 \Delta^* = \partial_r^2 + \frac{\sin \vartheta}{r^2} \partial_\vartheta \left( \frac{\partial_\vartheta}{\sin \vartheta} \right)
\end{equation}
denoting the Grad-Shafranov operator.\par
We choose $\Gamma_0=2.2877574$, $ \kappa = 65\text{ cm}^{2\Gamma_0-2}$ for the background system, being a polytrope similar to the analytical density profile \citet{Akguen} use. In order to quantify the impact of stratification on stability, we vary $\Gamma_1$ by choosing $\Delta \Gamma = \Gamma_1 - \Gamma_0$ between $0$ and \num{0.25}. We only consider stably stratified stars fulfilling $\Delta \Gamma >0$. That way, the investigated stars do not suffer from unstable g-modes in the unmagnetised state already. Investigating the stabilising impact of $B_{\text{pol}}$ on $B_{\text{tor}}$, the second parameter to vary is $\eta_{\text{pol}}$. The poloidal field strength is assumed to be always small compared to the toroidal one. The size of area $A$, where $\vec \xi \neq 0$ according to equation \eqref{equ:XiAkguen}, is adjusted for every $\Gamma_1$ by choosing
\begin{equation}
\label{equ:deltar}
 \delta_r = \num{2.617e-4} \delta_\vartheta \sqrt{\frac{\Gamma_0 \Gamma_1}{\Gamma_1 -\Gamma_0}} 
\end{equation}
accordingly. That way, the stability criterion analytically derived by \citet{Akguen} is kept valid. For the specific values used here, it reads
\begin{equation}
\label{equ:StabilityCriterionAkguen}
 \eta_{\text{pol}} > \eta_{\text{pol}}^{\text{crit}} =  \num{6.759e-5} \delta_\vartheta \sqrt{\frac{\Gamma_0 \Gamma_1}{\Gamma_1-\Gamma_0}}.
\end{equation}
This criterion is based on the assumption that both $\delta_r$ and $\delta_\vartheta$ approach zero simultaneously ($\delta_r \rightarrow 0$, $\delta_\vartheta \rightarrow 0$). Here we assume a small value $\delta_\vartheta = 0.05$, keeping the error of $\eta_{\text{pol}}^{\text{crit}}$ calculated via \eqref{equ:StabilityCriterionAkguen} small.

\subsubsection{Results}
\begin{figure}
	\includegraphics[width=\columnwidth]{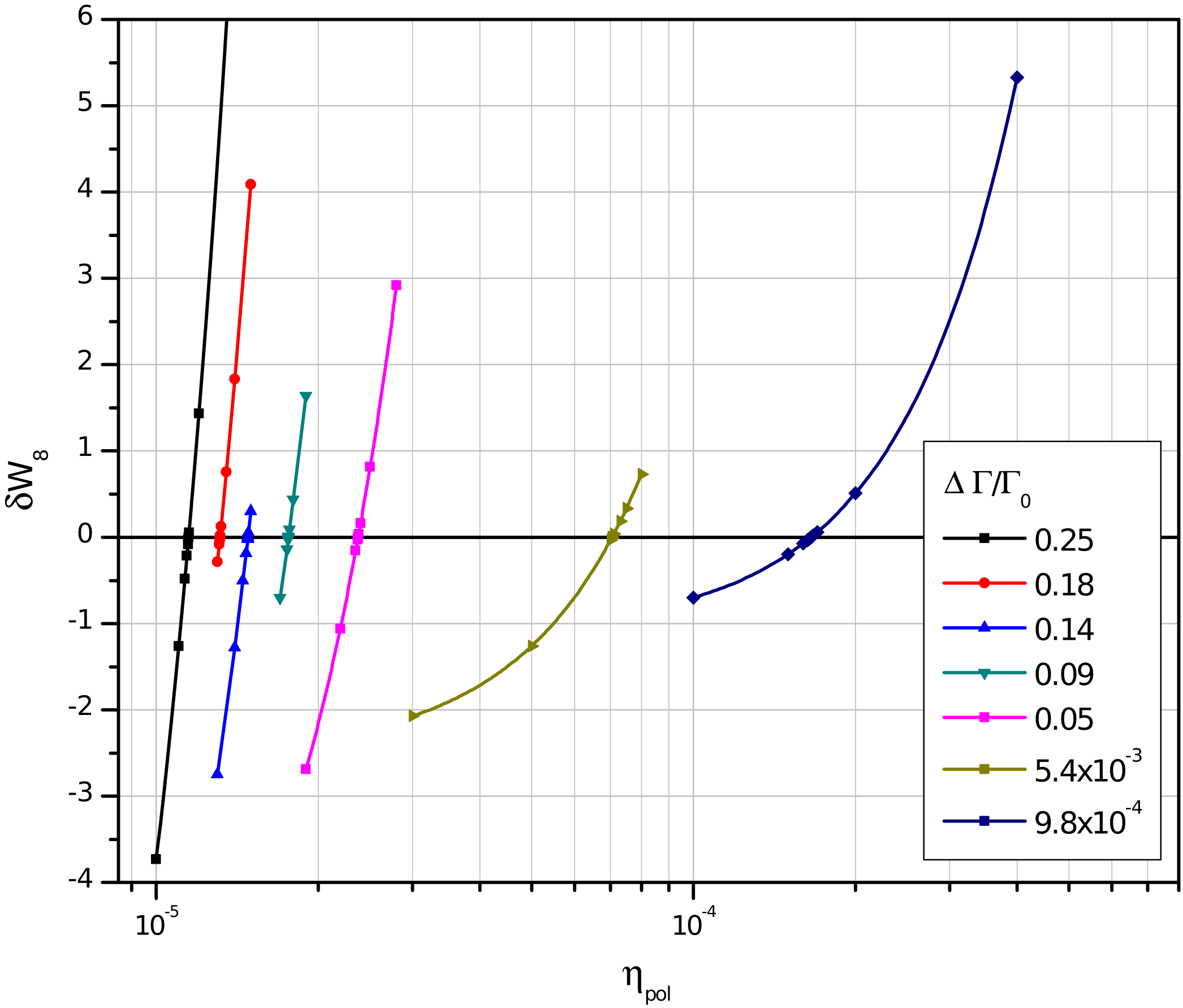}
    \caption{Calculated total energy variation in a mixed field system considering stratification in full treatment. $\delta W_8 \equiv \num{e8}\delta W$ is plotted against the maximum strength $\eta_\text{pol}$ of the poloidal component of the magnetic field for different levels of stratification $\Delta \Gamma/\Gamma_0$. All parameters being used are defined in section \ref{MixedFieldAndStratification}.}
\label{fig:deltaWvonnpol}
\end{figure}

The dependence of the calculated energy variation on the poloidal field strength in full non-Cowling treatment is given in Fig.~\ref{fig:deltaWvonnpol} for different degrees of stratification $\Delta \Gamma = \Gamma_1 - \Gamma_0$. The energy variation is negative for $\eta_{\text{pol}}=0$ for all $\Delta \Gamma$, implying that the system is unstable if the magnetic field is purely toroidal. The energy variation is positive if the poloidal field strength $\eta_{\text{pol}}$ exceeds a critical value, implying stabilisation of the system. This holds for every stably stratified star.\par
\begin{figure}
	\includegraphics[width=\columnwidth]{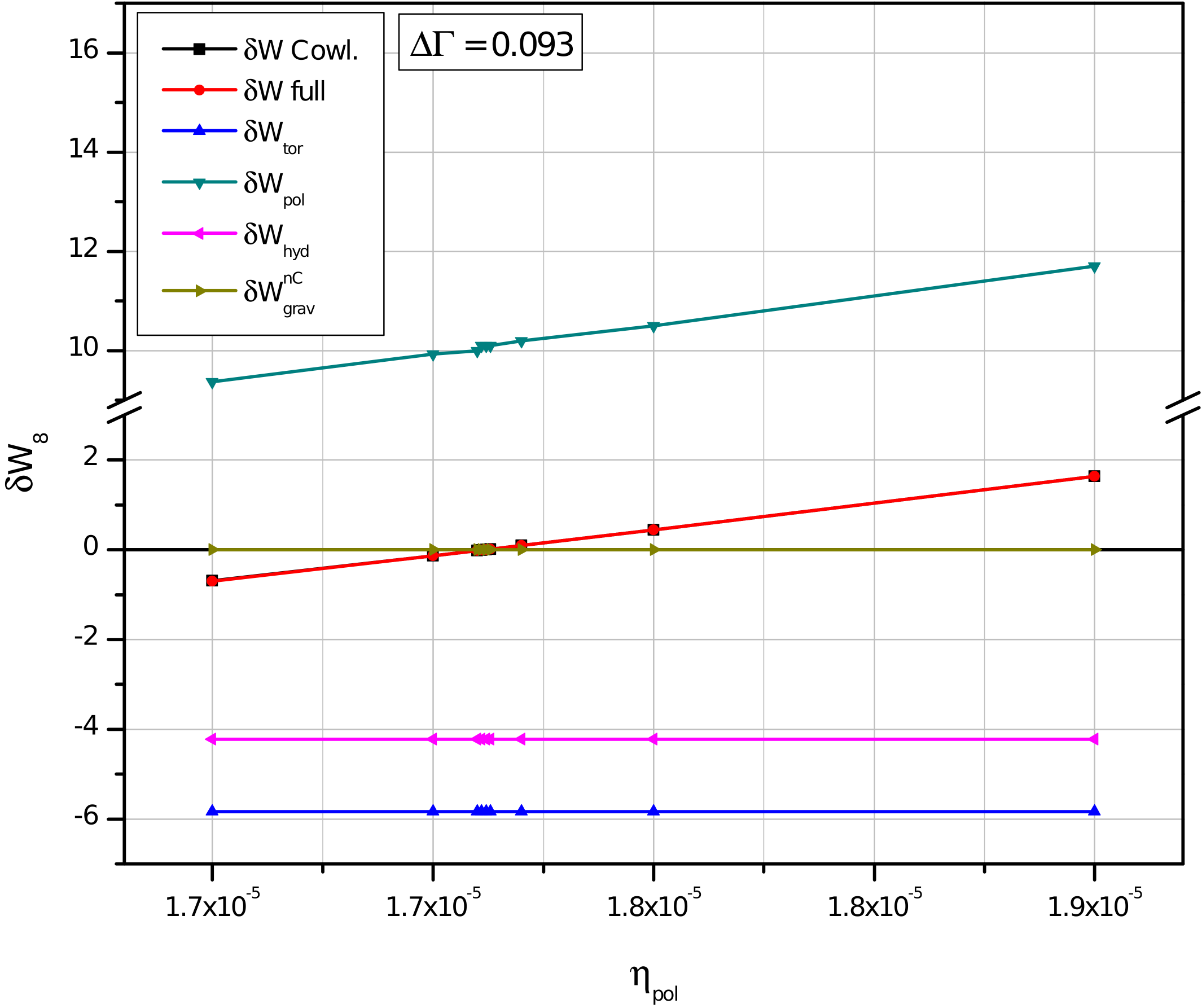}
    \caption{Calculated toroidal, poloidal, hydrostatic and non-Cowling contributions to the energy variation in a mixed field system with stratification. The plotted quantities are defined as $\delta W_8 \equiv \num{e8}\delta W$. The contribution of the non-Cowling gravity term $\delta W_{\text{grav}}^{\text{nC}}$ is negligibly small. Thus, the total energy variation in Cowling approximation and full description do not differ for the system under consideration. All parameters being used are specified in section \ref{MixedFieldAndStratification}.}
\label{fig:deltaWivonnpol}
\end{figure}

Fig.~\ref{fig:deltaWivonnpol} shows the separate contributions to the energy variation from magnetic field components, fluid pressure and gravity for an intermediate value of $\Delta \Gamma$. $\delta W_{\text{tor}}$ is negative, whereas $\delta W_{\text{pol}}>0$ grows with $\eta_{\text{pol}}$ as it was constructed. Hydrostatic contributions $\delta W_\text{hyd} = \delta W_{\text{fluid}} + \delta W_{\text{grav}}^{\text{Cowl}}$ and non-Cowling term $\delta W_{\text{grav}}^{\text{nC}}$, being neglected in \citet{Akguen}, are negative. $\left|\delta W_{\text{grav}}^{\text{nC}}\right|$ is negligibly small, there is almost no difference between the energy variation in Cowling approximation and in full treatment.
\par
\begin{figure}
	\includegraphics[width=\columnwidth]{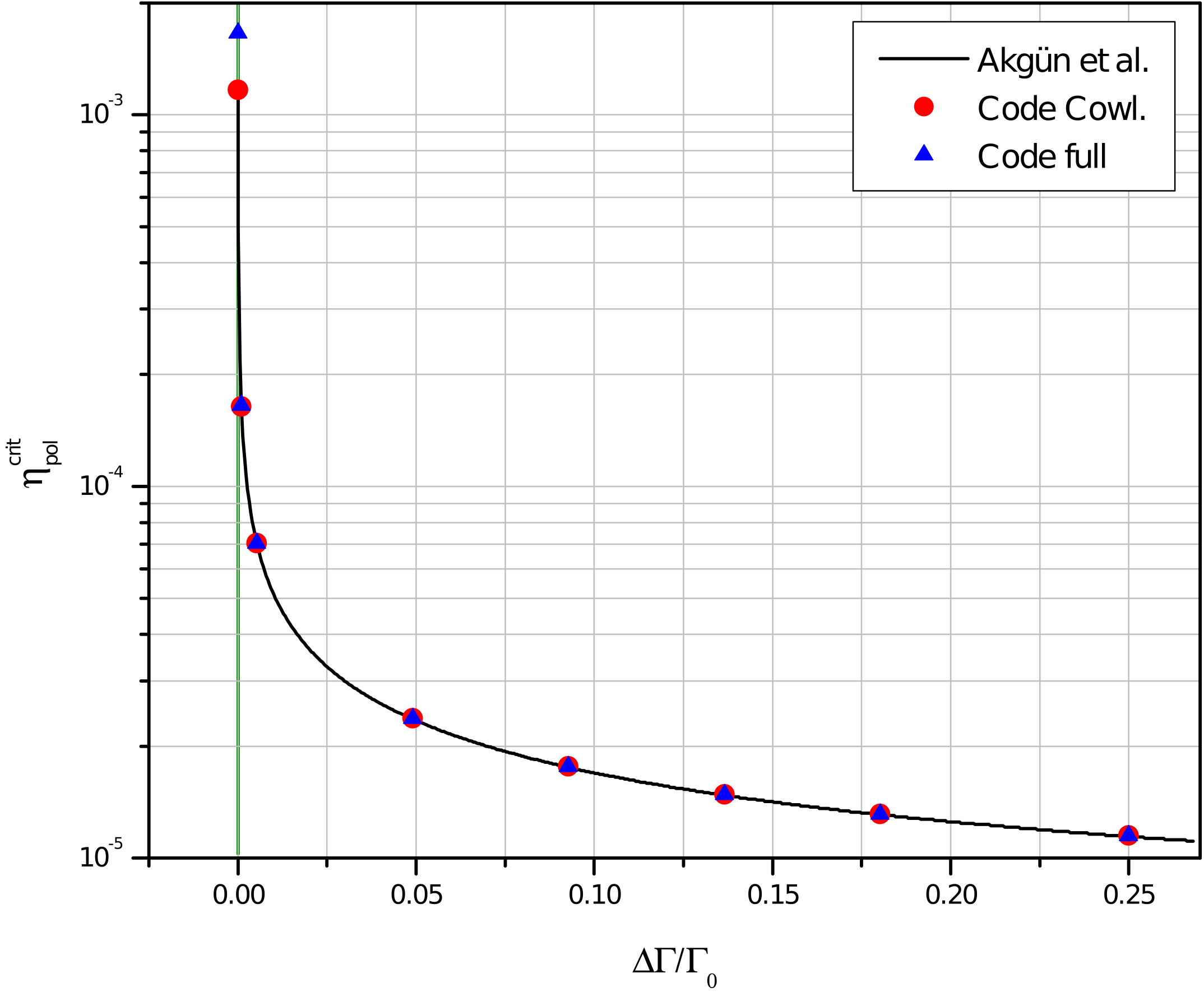}
    \caption{Critical poloidal field strength being necessary for stabilisation in a mixed field system considering stratification. The critical field strength $\eta_\text{pol}^\text{crit}$ is plotted against the level of stratification $\Delta \Gamma/\Gamma_0$. The calculated results in Cowling approximation and full treatment agree with the analytical stability criterion \eqref{equ:StabilityCriterionAkguen} from \citet{Akguen}.}
\label{fig:npolkritvonGamma1}
\end{figure}

\bigskip
The roots from the $\delta W(\eta_{\text{pol}})$-plot in Fig.~\ref{fig:deltaWvonnpol} give the critical poloidal field strength necessary for stability. They are plotted depending on the level of stratification in Fig.~\ref{fig:npolkritvonGamma1} for full and Cowling treatment, together with the analytical stability criterion \eqref{equ:StabilityCriterionAkguen}. Semi-analytic and analytical result are in great accordance. Remember that $\eta_{\text{pol}}$ and $\eta_{\text{tor}}$ define the strength of the poloidal and toroidal magnetic field components at their spatial maxima. For $\Gamma_1 \approx \Gamma_0$ the curve plotted according to the analytical criterion diverges, implying that no stabilisation is possible for any poloidal field strength if the star is not stratified. In the semi-analytic result, we cannot find an infinite value for the field strength due to the usage of numerical methods. Still, the calculated critical field strength exceeds the maximum field strength fulfilling $\eta_{\text{pol}} \ll \eta_{\text{tor}}$ by far, so that stabilisation is not possible for $\Gamma_0 = \Gamma_1$ according to the semi-analytic result as well. The strong agreement of both results also indicates that the slightly different choices that have been made for the background system here and in \citet{Akguen}
have basically no effect on the stability analysis. This comprises mainly the approximate analytical density distribution used in \citet{Akguen} versus the actual polytrope considered here. Still, the semi-analytic approach offers more flexibility than the analytical one by allowing for arbitrary choices of $\Gamma_0$ without re-derivation.
\par\bigskip
Comparison of the computed results between Cowling and full treatment in Fig.~\ref{fig:npolkritvonGamma1}, shows: The influence of $\delta W_{\text{grav}}^{\text{nC}}$ is negligible, even for $\Delta \Gamma \approx \num{0.25}$. In fact, the biggest deviation between both approaches is present for the smallest level of stratification we considered with $\Delta \Gamma = \num{1.86e-5}$. This deviation most probably does not represent a physical feature but lies within the range of numerical accuracy. The numerical error increases for $\Delta \Gamma \to 0$ at the transition between existing and non-existing g-modes. Altogether there is no significant difference between Cowling approximation and full treatment. This can be explained by the very small area $A$ with nonzero energy variational densities. Via relations \eqref{equ:deltawgravnC}, \eqref{equ:deltaPhiAnsatz} and \eqref{equ:deltaPhil}, $\delta W_{\text{grav}}^{\text{nC}}$ depends on $J_l$ and $K_l$, quantities being calculated by an integration over $r$, see expression \eqref{equ:JlKl2D}. In case of the extremely small radial extends $\delta_r = \num{4.4e-5}$ to $\num{9.5e-2}$ used here in \eqref{equ:deltar}, the effective region that contributes to the integrands of $J_l$ and $K_l$ is very small, in the analytical treatment infinitesimally small. As $d_r J_l$, $d_r K_l$ are not extremely steep, $J_l$, $K_l$ and subsequently $\delta W_{\text{grav}}^{\text{nC}}$ are small here.\par
One should mention though that the displacement field being considered here is rather constructed than natural in the following sense. Here we assume $|\vec \xi|$ to be a step function having no impact on the surroundings of $A$. A realistic displacement field on the other hand extends over the whole star due to couplings in the fluid around the perturbation centre.\par
The result from \citet{Akguen} shows in valid Cowling approximation how the Tayler instability of the toroidal field can be stabilised locally. For global stabilisation with spatially extended perturbation modes the consideration in full treatment may be relevant though. $\delta \Phi$ might also have a significant impact in a fully relativistic treatment where changes in the gravitational potential caused by a displacement behave differently than in a simplified Newtonian framework.

\section{Conclusions}
\label{Conclusions}

We constructed a semi-analytic stability analysis scheme based on the energy variational principle. It requires less simplifications than analytical treatments and has the ability to receive general information contrary to pure numerical simulations. We applied it to different systems of magnetised neutron stars, addressing the unresolved question of the interior magnetic field structure.\par\bigskip
First applications show on one side that the semi-analytic approach works and provides correct results that are in accordance with known analytical results: We verified Tayler instabilities for purely toroidal and poloidal fields in sections \ref{PurelyToroidalMagneticField} and \ref{PurelyPoloidalMagneticField} and the results for mixed fields and stratification of \citet{Akguen} in section \ref{MixedFieldAndStratification}. These results have been achieved without noticeable accuracy loss compared to fully analytical studies.\par
On the other side, the idea of the semi-analytic approach works, providing advantages in stability analysing magnetised neutron stars. Compared to an analytical treatment we were able to drop common simplifications: We don't rely on using Cowling approximation anymore, the semi-analytic scheme allowed us to implement stratification in the considered system without complications and to use arbitrary background configurations, such as polytropes. These improvements towards a pure analytical procedure have been applied in section \ref{MixedFieldAndStratification}.\par
Along with this, we set up an analytical expression for the Eulerian perturbation of the gravitational potential in an axisymmetric system.\par\bigskip
Further advantages of our method lead to possible future applications.\par
Plotting the energy variational integrand $\mathcal E$ depending on spatial variables or system parameters directly yields a global understanding of the problem under consideration. This is a possibility neither an analytical nor numerical approach can offer that easily.\par
The most promising application of the scheme lies in searching for stable equilibria in magnetised neutron stars. Making the system more realistic by adding characteristics like the neutron star crust can give us information about the relevance different effects have in stabilising magnetised neutron stars. Plotting the energy variation $\delta W$ depending on system parameters may show us `isles' of stability for certain parameter ranges and corresponding composition or magnetic field configurations.\par
Another advantage of the semi-analytic approach lies in simplifying the attempt of searching for stability rather than instabilities. In contrast to earlier studies of simple models \citep{Wright, Tayler1, Tayler2, Tayler3}, latest works \citep{BraithwaiteA, LanderJones2012, Akguen} focus on explaining the observed stability of magnetised neutron stars and resolving the instability issue rather than finding instabilities. Proving instability requires finding one unstable displacement field, whereas stability detection involves testing the system stability against all possible displacement fields, cf. stability criterion \eqref{equ:StabilityCriterion}. Addressing this far more complex task, we suggest a systematic construction of displacement fields expressed in the full set of stellar eigenfunctions \citep{Unno, Fluegge, Smeyers}. This assumption can easily be implemented within the semi-analytic scheme as opposed to a pure analytical approach.

\section*{Acknowledgements}
We thank the CARL-ZEISS foundation for financial support of this work. Our thanks also go to Taner Akg\"{u}n, Simone Dall'Osso, Sam Lander, Alpha Mastrano, Andrew Melatos and Pantelis Pnigouras for helpful discussions and inspiring ideas.


\bibliographystyle{mnras}
\bibliography{Literatur}


\appendix

\section{Derivations}
\subsection{Derivation of the energy variation}
\label{DerivationEnergyVariation}
In order to receive the explicit expression for the energy variational density \eqref{equ:deltawi} from the general form \eqref{equ:deltawiZwischenschritt}, all Eulerian perturbations have to be calculated perturbing the corresponding equations \citep{Mestel, Bernstein}
\begin{subequations}
 \label{equ:PerturbedEquations}
\begin{align}
 \vec j_0 +\delta \vec j &= \vec \nabla \times \vec B_0 + \vec \nabla \times \delta \vec B + \mathcal O\left(\delta^2\right)\\
 \partial_t \left( \vec B_0 + \delta \vec B \right) &= \vec \nabla \times \left( \dot{\vec \xi} \times \vec B_0 \right) + \vec \nabla \times \left( \vec v_0 \times \vec B_0 \right) + \mathcal O \left(\delta^2 \right)\\
 \partial_t \left( \rho_0 + \delta \rho \right) &= - \vec \nabla \cdot  \left( \rho_0 \vec v_0 \right) - \vec \nabla \cdot  \left( \rho_0 \dot{\vec \xi} \right) - \vec \nabla \cdot  \left( \delta \rho \vec v_0 \right) +\mathcal O\left(\delta^2\right)\\
 \label{equ:PerturbedPoissonEquation}
  \vec \nabla \cdot \vec \nabla \left( \Phi_0 + \delta \Phi \right) &= \mp 4 \pi \left(\rho_0 + \delta \rho \right) +\mathcal O\left(\delta^2\right)
\end{align}
\end{subequations}
using the equilibrium equations
\begin{subequations}
 \label{equ:EquilibriumEquations}
\begin{align}
 \vec j_0 &=  \vec \nabla \times \vec B_0\\
 \partial_t \vec B_0 &= \vec \nabla \times \left( \vec v_0  \times \vec B_0 \right)\\
\partial_t \rho_0 &= - \vec \nabla \cdot  \bigl( \rho_0 \vec v_0 \bigr)\\
\label{equ:EquilibriumPoissonEquation}
 \vec \nabla \cdot \vec \nabla \delta \Phi_0 &= 4 \pi \rho_0.
\end{align}
\end{subequations}
$\delta \vec j$ as given by \eqref{equ:deltaj}, $\delta \vec B$ shown in \eqref{equ:deltaB} and $\delta \rho$ given in \eqref{equ:deltarho} follow immediately from equations \eqref{equ:PerturbedEquations} combined with expressions \eqref{equ:EquilibriumEquations} under the assumption of fluid movements caused by the perturbation being fast compared to fluid movements in equilibrium, $\dot {\vec \xi}=\delta \vec v \gg \vec v_0$. In the case of magnetic field and density, one time integration has been carried out.\par
Using equilibrium equation \eqref{equ:EquilibriumPoissonEquation} and $\delta \rho$ from \eqref{equ:deltarho} in the perturbed Poisson equation \eqref{equ:PerturbedPoissonEquation}, yields
\begin{equation}
 \vec \nabla \cdot \vec \nabla \delta \Phi = \pm 4 \pi \delta \rho =\pm 4 \pi \vec \nabla \cdot \left( \rho_0 \vec \xi \right)
\end{equation}
for the Eulerian perturbation of the gravitational potential, which is of the form of a Poisson equation itself. For a known distribution of $\vec \nabla \cdot \left( \rho_0 \vec \xi \right)$ it is formally solved by
\begin{equation}
 \delta \Phi(\vec r) = \pm G \int \frac{ \vec \nabla \cdot \left[\rho_0(\vec r') \vec \xi (\vec r') \right]}{|\vec r - \vec r'|} \d V'.
\end{equation}
Applying partial integration in three dimensions,
\begin{equation}
 \int\limits_V U \vec \nabla \cdot \vec V \d V = \oint\limits_{\partial V} U \vec V \d \vec S - \int\limits_V \vec V \cdot \vec \nabla U \d V, 
\end{equation}
with $U=\frac{1}{|\vec r-\vec r'|}$, $\vec V = \rho_0(\vec r') \vec \xi(\vec r')$ and taking into account that the surface integral vanishes due to $\rho_0(R)=0$, expression \eqref{equ:deltaPhi} for $\delta \Phi$ follows as given in \citep{ChandrasekharLebovitz}.\par
The Eulerian pressure perturbation can be written as
\begin{equation}
\delta p =  \left(\frac{\partial p_0}{\partial \rho_0}\right)_{x^p} \delta \rho + \left(\frac{\partial p_0}{\partial x^p_0}\right)_\rho \delta x^p
					= \frac{p_0 \Gamma_1}{\rho_0} \delta \rho + \frac{p_0 \Gamma_p}{x^p_0} \delta x^p,
\end{equation}
where
\begin{equation}
\Gamma_1 =  \left(\frac{\partial \ln p_0}{\partial \ln \rho_0}\right)_{x^p} = \frac{p_0}{\rho_0}  \left(\frac{\partial p_0}{\partial \rho_0}\right)_{x^p}, \quad 
\Gamma_p \equiv  \left(\frac{\partial \ln p_0}{\partial \ln x^p_0}\right)_{\rho} = \frac{p_0}{x^p_0}  \left(\frac{\partial p_0}{\partial x^p_0}\right)_{\rho}
\end{equation}
have been used. Next, $\delta x^p = - \vec \xi \cdot \vec \nabla x_0^p$, which follows from $\Delta x^p =0$ and relation \eqref{equ:DeltaQdeltaQ}, as well as $\delta \rho$ from \eqref{equ:deltarho} are inserted to give
\begin{equation}
\label{equ:deltapZwischenschritt}
 \delta p = -\frac{p_0 \Gamma_1}{\rho_0} \vec \nabla \cdot \left(\rho_0 \vec \xi \right) - \frac{p_0 \Gamma_p}{\rho_0} \frac{\d \ln x^p_0}{\d \ln \rho_0} \vec \xi \vec \nabla  \rho_0,
\end{equation}
where the expression
\begin{equation}
 \vec \nabla  x^p_0 = \frac{\d x^p_0}{\d \rho_0} \vec \nabla  \rho_0 = \frac{x^p_0}{\rho_0} \frac{\d \ln x^p_0}{\d \ln \rho_0} \vec \nabla  \rho_0
\end{equation}
has been used. From the dependence $p=p(\rho,x^p)$ in the stratified star (see section \ref{Stratification}), the equilibrium fluid pressure gradient can be written as
\begin{align}
 \vec \nabla p_0 &= \left(\frac{\partial p_0}{\partial \rho_0}\right)_{x^p} \vec \nabla \rho_0 + \left(\frac{\partial p_0}{\partial x^p_0}\right)_\rho \frac{\d x^p_0}{\d \rho_0} \vec \nabla \rho_0\\
 \label{equ:gradp0gradrho0}
		&= \frac{p_0}{\rho_0} \underbrace{\left[ \left( \frac{\partial \ln p_0}{\partial \ln \rho_0}\right)_{x^p} +  \left( \frac{\partial \ln p_0}{\partial \ln x^p_0} \right)_\rho \frac{\d \ln x^p_0}{\d \ln \rho_0} \right]}_{\Gamma_0} \vec \nabla \rho_0,
\end{align}
where the background polytropic exponent has been identified via its definition \eqref{equ:Gamma0}. Using relation
\begin{equation}
\Gamma_p \frac{\d \ln x^p_0}{\d \ln \rho_0} = \Gamma_0 - \Gamma_1
\end{equation}
from \citep{Passamontilang} and applying relation \eqref{equ:gradp0gradrho0} in expression \eqref{equ:deltapZwischenschritt}, the form of $\delta p$ given in \eqref{equ:deltap} finally follows.

\subsection{Rewriting of hydrostatic energy variational density}
\label{EquivalenceWithAkguen}
According to expression \eqref{equ:deltawi}, the hydrostatic contribution to the energy variational density in Cowling approximation in this work is
\begin{align}
\label{equ:deltawhydZwischenschritt}
\mathcal E_{\text{fluid}} +  \mathcal E^{\text{Cowl}}_{\text{grav}} = \frac{1}{2} &\Re \left[ \Gamma_1 p_0 \left( \vec \nabla \cdot \vec \xi^* \right)\left( \vec \nabla \cdot \vec \xi \right)  \right.\\
\nonumber & \left. + \left( \vec \xi^* \cdot \vec \nabla p_0 \right) \vec \nabla \cdot \vec \xi \mp \left( \vec \xi^* \cdot \vec \nabla \Phi_0 \right) \vec \nabla\cdot \left(\rho_0 \vec \xi\right) \right].
\end{align}
With relation \eqref{equ:gradp0gradrho0}, the hydrostatic equilibrium equation \eqref{equ:HydrostaticEquilibrium} yields $\vec \nabla \Phi_0 = \pm \frac{\Gamma_0 p_0}{\rho_0^2} \vec \nabla \rho_0$. Inserted into \eqref{equ:deltawhydZwischenschritt} and using $\vec \xi \cdot \vec \nabla U = \vec \nabla \cdot \left( U \vec \xi \right) - \vec \xi \cdot \vec \nabla U$,
\begin{equation}
\mathcal E_{\text{fluid}} +  \mathcal E^{\text{Cowl}}_{\text{grav}} = \frac{1}{2} \Re \left[ (\Gamma_1 - \Gamma_0) p_0 \left(\vec \nabla \cdot \vec \xi \right)^2 + \frac{\Gamma_0 p_0}{\rho_0^2} \left(\vec \nabla \cdot \left(\rho_0 \vec \xi \right)\right)^2 \right]
\end{equation}
follows, being the hydrostatic part of the energy variational density given in \citet{Akguen}.

\section{Numerical procedure in detail}
\label{NumericalProcedureInDetail}
We calculate the energy variation by integrating the energy variational density given by expression \eqref{equ:deltawi}. This procedure has been summed up in section \ref{NumericalProcedure}. Here we describe the numerical part of our work in more detail.

\subsection{Analytical preparation of the integral}
\label{AnalyticalPreparationOfTheIntegral}
Due to the axisymmetry of the model system, $Q_0$ do not depend on the azimuthal angle $\varphi$. All of the choices \eqref{equ:XiTaylerToroidal}, \eqref{equ:XiTaylerPoloidal} and \eqref{equ:XiQuerAkguen} we use for the displacement field, Fourier-analyse the $\varphi$-dependence of $\vec \xi$. That way, the $\varphi$-integration in \eqref{equ:deltaW} can be carried out analytically, reducing the numerical effort to two-dimensional integration. The numerical integration is carried out in cylindrical coordinates $(\varpi,z)$ being best adjusted to the magnetic field geometry.\par
The $\vec \xi$-defining functions $X$, $Y$ and $Z$ as well as $\tilde R$, $\tilde S$ and $\tilde T$ are chosen as functions of $\varpi$, $z$, equilibrium quantities and parameters, see assumptions \eqref{equ:ChoiceTaylerToroidalXZ}, \eqref{equ:ChoiceTaylerPoloidalXYZ} and \eqref{equ:ChoiceAkguenRS}. Thus, the real parts in the integrand \eqref{equ:deltawi} can be evaluated. The remaining explicit integrands given by \eqref{equ:deltaWexplicitBtorB}, \eqref{equ:deltaWexplicitBpol} and \eqref{equ:deltaWexplicitAkguen} depend on position variables, equilibrium quantities and parameters only. They are evaluated numerically.

\subsection{Calculation of equilibrium quantities}
\label{CalculationOfEquilibriumQuantities}
\begin{figure}
	\includegraphics[width=0.49\columnwidth]{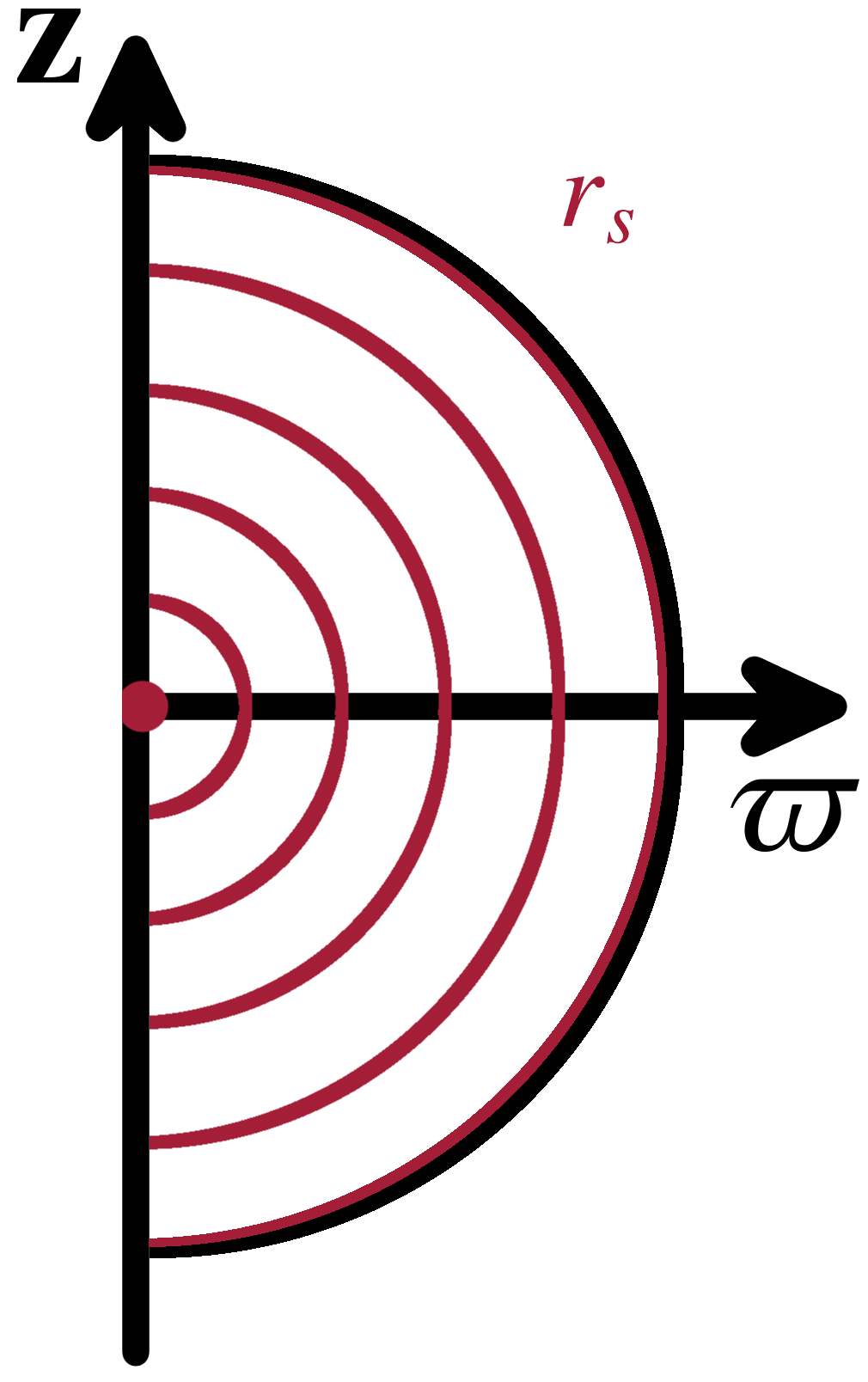}
	\includegraphics[width=0.47\columnwidth]{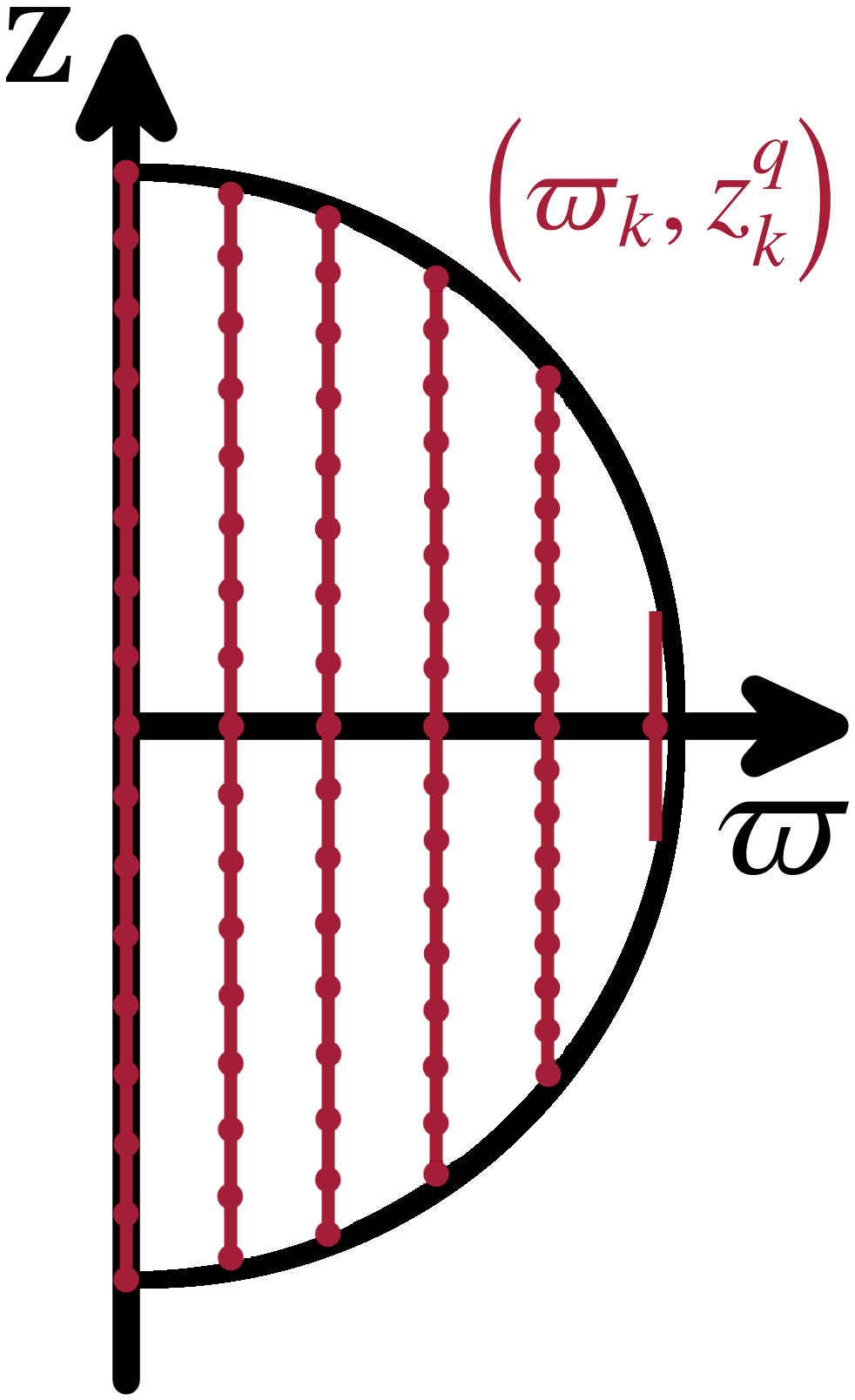}
    \caption{Grid points $r_s$ for the calculation of background quantities in spherical symmetry (left); grid points $(\varpi_k, z_q)$ for the integration of the energy variational density \eqref{equ:deltaW} in cylindrical coordinates, calculated with splines interpolation (right).}
\label{fig:IntegrationGrid}
\end{figure}

In order to calculate the hydrostatic equilibrium quantities appearing in the integrands of \eqref{equ:deltaWexplicitBtorB}, \eqref{equ:deltaWexplicitBpol} and \eqref{equ:deltaWexplicitAkguen}, we solve the system equations \eqref{equ:SystemEquations} yielding $m_0(r)$, $p_0(r)$, $\rho_0(r)$, $\Phi_0(r)$. For this, we apply a classical Runge--Kutta method under the assumption of a polytropic equation of state \eqref{equ:EquationOfStateBackground}. In a second step, we make use of cubic splines to interpolate the quantities from radial grid points $r_s$ to the required cylindrical integration grid points $(\varpi_k, z_q)$ as shown in Fig.~\ref{fig:IntegrationGrid}, yielding $m_0(\varpi,z)$, $p_0(\varpi, z)$, $\rho_0(\varpi,z)$, $\Phi_0(\varpi,z)$.\par
Note that the semi-analytic approach is not constrained on using a polytropic equation of state nor approximating the equilibrium quantities by that of an unmagnetised star. Neither must the background system fulfil spherical symmetry. The fact that the integration is being carried out numerically also allows for a numerical solution of the system equations. In principle, any arbitrary data set for axisymmetric equilibrium quantities can be used.\par
The equilibrium magnetic field, analytically defined in a parametrized form according to \eqref{equ:BTaylerToroidal}, \eqref{equ:BTaylerPoloidal} or \eqref{equ:BAkguen}, can directly be used in the numerical integrand.

\subsection{Calculation of the non-Cowling term}
In case of the full treatment without Cowling approximation, $\delta \Phi$ including $J_l$ and $K_l$ appears in the energy variational integrand \eqref{equ:deltawi}. It has to be calculated by solving the two-dimensional integrals \eqref{equ:JlKl2D} defining $J_l$ and $K_l$ before the actual $\mathcal E$-integration can be carried out.\par
In this work, we consider the full treatment in the case of \emph{mixed magnetic field with stratification} where $\vec \xi$ is given by expression \eqref{equ:XiAkguen} with \eqref{equ:XiQuerAkguen}. Inserting this assumption into \eqref{equ:JlKl2D} and making use of the routine providing the equilibrium quantities we described in \ref{CalculationOfEquilibriumQuantities}, the integrands are fully determined and can be computed numerically. As $J_l$ and $K_l$ do not depend on the magnetic field, the integration is performed in spherical coordinates $(r', \vartheta')$.

\subsection{Integration area}
\label{IntegrationArea}
The calculation of the total energy variation requires the integration area of \eqref{equ:deltaW} to extend over the stellar volume $V$. Local instabilities enclosed within subareas of the star can be made visible by choosing smaller integration areas.\par
Showing Tayler's instability in the case of a \emph{purely toroidal} field with assumptions \eqref{equ:BTaylerToroidal} for $\vec B$ and \eqref{equ:XiTaylerToroidal} for $\vec \xi$, we choose the integration area to be equal to $V$. Thus, the 2D-integration bounds are
\begin{equation}
\label{equ:IntegrationAreaV}
 \begin{array}{l}
  \varpi_a = 0\\
  \varpi_b =R
 \end{array}
 \qquad \qquad
  \begin{array}{l}
  z_a = -\sqrt{R^2 - \varpi^2}\\
  z_b = +\sqrt{R^2 - \varpi^2},
 \end{array}
\end{equation}
with $a$ and $b$ denoting lower and upper bound. This 2D-integration area corresponds to half of the stellar cross section $S(\text{star})$ in the $(\varpi, z)$-plane for constant $\varphi$.\par
For the \emph{purely poloidal} magnetic field we assume \eqref{equ:BTaylerPoloidal} for $\vec B$ and \eqref{equ:XiTaylerPoloidal} for $\vec \xi$. Different from the other cases, field lines penetrating the stellar surface do exist here. When integrating over $V$, they lie partly outside the integration area and make a positive contribution to the energy variation. This contribution can cover local instabilities which contribute small negative summands to $\delta W$. These instabilities might occur in the closed field line region around the magnetic axis we are interested in. In order to avoid this, we exclude the open field line region around the symmetry axis from the integration area. For an appropriate choice of $R_{\text{tor}}$ and $r_{\text{tor}}$, the torus on which the toroidal coordinate system is defined according to expression \eqref{equ:ToroidalCoordinates} is a suitable integration area. The 2D-integration bounds are
\begin{equation}
\label{equ:IntegrationAreaTorus}
  \begin{array}{l}
  \varpi_a = R_{\text{tor}}-r_{\text{tor}}\\
  \varpi_b = R_{\text{tor}}+r_{\text{tor}}
 \end{array}
 \qquad
  \begin{array}{l}
  z_a = -\sqrt{r_{\text{tor}}^2 - (R_{\text{tor}}-\varpi)^2}\\
  z_b = +\sqrt{r_{\text{tor}}^2 - (R_{\text{tor}}-\varpi)^2}.
 \end{array}
\end{equation}
This area $S(\text{torus})$ corresponds to the cross section of the torus for constant $\varphi$.\par
\begin{figure}
	\includegraphics[width=\columnwidth]{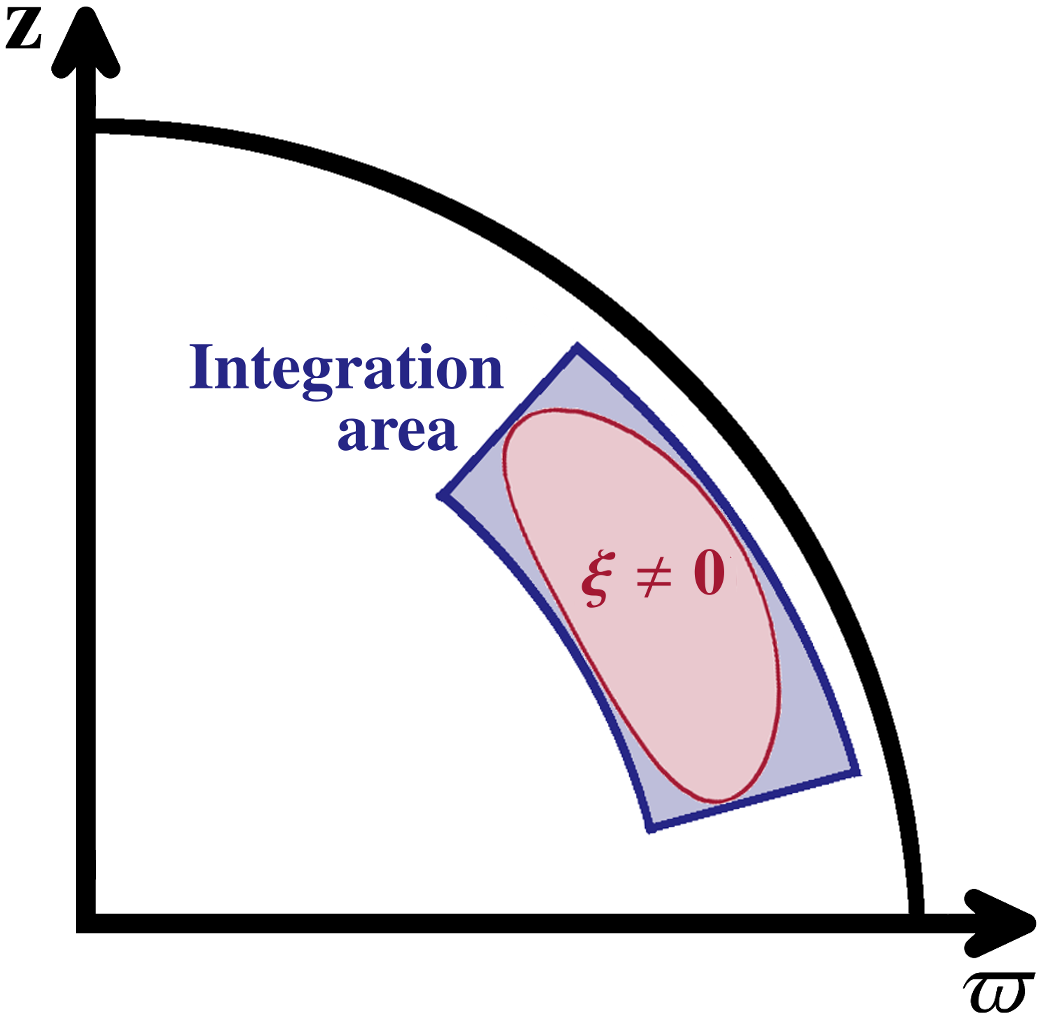}
    \caption{Integration area chosen for calculating $J_l$ and $K_l$, as given in \eqref{equ:JlKl2D} in the case of mixed fields with stratification. The integration area includes area $A$, where $\vec \xi \neq 0$ according to \eqref{equ:XiAkguen}. The illustration of both areas is strongly enlarged compared to the actual sizes being used for computation.}
\label{fig:IntegrationAreaAkguen}
\end{figure}

In the case of \emph{mixed magnetic fields} given by \eqref{equ:BAkguen} \emph{and stratification}, the nonzero displacement field as chosen by \eqref{equ:XiAkguen} is strongly confined to the area $A$. The energy variational integrand $\mathcal E=\mathcal E(\vec \xi)$ vanishes everywhere outside $A$, by what integration over $V$ and $A$ become equivalent. In calculating $\delta W$, one two-dimensional integration has to be performed for every given set of parameters. The integral contributions of most grid points vanish such that computing time and power are acceptable even if the stellar volume is being used as an integration area. Thus, we keep the integration area simple by choosing $S(\text{star})$ as defined in \eqref{equ:IntegrationAreaV} here as well. On the contrary, in calculating $J_l$ and $K_l$ as defined in \eqref{equ:JlKl2D}, a two-dimensional integration is necessary for every $r_s$-grid point that is used for the $\delta W$-calculation. For the sake of computing power and numerical accuracy, we use an integration area not much bigger than $A$ in this case. The 2D-integration bounds are
\begin{equation}
  \begin{array}{l}
  r'_a = (x_0-\delta_r) R\\
  r'_b = (x_0 +\delta_r) R
 \end{array}
 \qquad
  \begin{array}{l}
  \vartheta'_a = \vartheta_0-\delta_\vartheta\\
  \vartheta'_b = \vartheta_0+\delta_\vartheta.
 \end{array}
\end{equation}
This area has a rectangular shape in the $(r, \vartheta)$-plane. Its side lengths equal the maximum extents of $A$ along $r$ and $\vartheta$ , see Fig.~\ref{fig:IntegrationAreaAkguen}.

\subsection{Numerical integration scheme}
After setting up the integrands in a numerically amenable form, the integration is carried out.\par
In case of a full non-Cowling treatment, $J_l(r_s)$ and $K_l(r_s)$ are calculated by integrating \eqref{equ:JlKl2D} using Simpson's method twice along $\vartheta'$ and $r'$ first. Next, cubic splines are used to interpolate $J_l(r_s)$ and $K_l(r_s)$ to the cylindrical grid points $(\varpi_k, z_q)$. That way, $\delta \Phi$ as given by \eqref{equ:deltaPhi} can be expressed explicitly in the energy variational integrand \eqref{equ:deltawi}.\par
Finally, $\delta W$ is computed performing two Simpson integrations in \eqref{equ:deltaWexplicitBtorB}, \eqref{equ:deltaWexplicitBpol} and \eqref{equ:deltaWexplicitAkguen} along $\varpi$ and $z$ respectively.



\bsp	
\label{lastpage}
\end{document}